\documentclass[prc,aps,preprint,showpacs,nofootinbib,showkeys,eqsecnum,tightenlines]{revtex4}    
\usepackage{epsfig}
\def\be{\begin{equation}}
\def\ee{\end{equation}}
\def\bea{\begin{eqnarray}}
\def\eea{\end{eqnarray}}
\def\ot{(1\,\leftrightarrow\,2)}
\def\sa{\vec \sigma_1}
\def\sb{\vec \sigma_2}
\def\qa{\vec q_1}
\def\qb{\vec q_2}
\def\vq{\vec q}
\def\P{\vec P}
\def\Pa{\vec P_1}

\def\s2q2{(\vec \sigma_2 \cdot \vec q_2)}
\def\ta{\,\tau^{\,a}_1}
\def\tba{\,\tau^{\,a}_2}
\def\tta{\,i\,(\vec \tau_1 \times \vec \tau_2)^a}
\def\fot{\frac{1}{2}}
\def\vep{\varepsilon}
\def\st{\sqrt{2}}
\def\trejzero#1#2#3{
\left( \begin{array}{ccc}
                     #1 &#2 &#3 \\
                      0 & 0 & 0
       \end{array}
\right)
}
\def\trej#1#2#3#4#5#6{
\left( \begin{array}{ccc}
                     #1   &#2  &#3   \\
                     #4   &#5  &#6
       \end{array}
\right)
}
\def\seij#1#2#3#4#5#6{
\left\{ \begin{array}{ccc}
                     #1   &#2  &#3   \\
                     #4   &#5  &#6
        \end{array}
\right\}
}
\def\novej#1#2#3#4#5#6#7#8#9{
\left\{ \begin{array}{ccc}
                     #1   &#2  &#3   \\
                     #4   &#5  &#6   \\
                     #7   &#8  &#9
        \end{array}
\right\}
}
\newcommand{\noi}{\noindent}

\def\Journal#1#2#3#4{{#1}{\bf #2} (#4) #3 }
\def\NPA{{ Nucl. Phys.} \bf A}
\def\NPB{{ Nucl. Phys.} \bf B}
\def\PRL{ Phys. Rev. Lett.\,\,}
\def\PRC{{ Phys. Rev.}  C}
\def\PRD{{ Phys. Rev.}  D}

\def\FBS{ Few--Body Systems\,\,}
\def\PLB{{ Phys. Lett.} \bf B}

\def\EPJA{{ Eur. Phys. J.}\bf A}

\def\ZPC{{ Z. Phys.} \bf C}
\def\PR{ Phys. Rep.\,\,}
\def\IJMPA{{ Int. J. Mod. Phys.} \bf A}
\def\IJMPE{{ Int. J. Mod. Phys.} \bf E}

\def\HI{ Hyperfine Interactions\,\,}
\def\SJPN{ Sov. J. Part. Nucl.\,\,}
\def\APNY{ Ann. Phys. (N.Y.)\,\,}

\def\RMP{ Rev. Mod. Phys.\,\,}
\def\ANP{ Adv. Nucl. Phys.\,\,}
\def\JPG{ J. Phys. \bf G}

\def\CJPB{{ Czech. J. Phys.} \bf B}
\def\PPNP{Prog. Part. Nucl. Phys.\,\,}

\def\HJS{Hadronic J. Suppl.\,\,}

\begin{document}
%
%
\title{
{\large{\bf Muon Capture in Deuterium}}}
\author{P. Ricci}
\affiliation{Istituto Nazionale di Fisica Nucleare, Sezione di
Firenze, I-50019, Sesto Fiorentino (Firenze), Italy }
\author{E.~Truhl\'{\i}k}
\affiliation{Institute of Nuclear Physics ASCR, CZ--250 68
\v{R}e\v{z}, Czech Republic }
\author{B. Mosconi}
\affiliation{Universit$\grave{a}$ di Firenze, Department of Physics,
and Istituto Nazionale di Fisica Nucleare, Sezione di Firenze,
I-50019, Sesto Fiorentino (Firenze), Italy }
\author{J. Smejkal}
\affiliation{Institute of Technical and Experimental Physics, Czech
Technical University, CZ--128 00 Horsk\'a 3a/22, Prague, Czech
Republic }

\begin{abstract}

Model dependence of the capture rates of the negative muon capture
in deuterium is studied starting from potential models and the weak
two-body meson exchange currents constructed in the tree
approximation and also from an effective field theory. The tree
one-boson exchange currents are derived from the hard pion chiral
Lagrangians of the $N \Delta \pi \rho \omega a_1$ system. If
constructed in conjunction with the one-boson exchange potentials,
the capture rates can be calculated consistently. On the other hand,
the effective field theory currents, constructed within the heavy
baryon chiral perturbation theory, contain a low energy constant
$\hat d ^R$ that cannot be extracted from data at the one-particle
level nor determined from the first principles. Comparative analysis
of the results for the doublet transition rate allows us to extract
the constant $\hat d ^R$.

\end{abstract}

\noi \pacs{12.39.Fe; 21.45.Bc; 23.40.-s}

\noi \hskip 1.9cm \keywords{negative muon capture; deuteron;
potential models; effective field theory; meson exchange currents}

\maketitle

\section{Introduction}
\label{intro}

The study of the weak interaction in the deuterium at low energies
is an important topic. Since the structure of the nucleon-nucleon
(NN) force is at present well understood and the weak vector one-
and two-nucleon interactions and the weak axial one-nucleon
interaction are also well known, it can provide the most reliable
information on the structure of the space component of the weak
axial meson exchange currents (MECs). This component contributes
notably into the capture rates for the reaction of the negative muon
capture in deuterium \be
\mu^-\,+\,d\,\longrightarrow\,\nu_\mu\,+\,n\,+\,n\,, \label{MUD} \ee
from the hyperfine $\mu d$ states.

In reaction (\ref{MUD}), the stopped muons are captured from the
doublet or quadruplet hyperfine states with total angular momentum
F=1/2 or 3/2, respectively. The corresponding capture rates,
$\Lambda_{1/2}$ and $\Lambda_{3/2}$, were calculated in the past by
many authors (for the references, see \cite{KIT,DFM}). Since
$\Lambda_{3/2}\,\approx\,10$ s$^{-1}$ and
$\Lambda_{1/2}\,\approx\,400$ s$^{-1}$, only the doublet capture
rate is of practical interest.

Until recently, the capture rates were calculated with the deuteron
and neutron-neutron wave functions derived from the variety of
nuclear potentials of the first generation and with the MECs
constructed in the tree approximation. The systematic investigation
of the structure of the operator of the tree weak axial MECs based
on the chiral Lagrangians was started in Ref.\,\cite{IT1}. These
MECs were applied in calculations of the capture rates and
neutron-neutron spectra for reaction (\ref{MUD})in
Refs.\,\cite{IT2,DET,ATCS,TKK}. The obtained values of the doublet
capture rate $\Lambda_{1/2}$ were in the interval 398
s$^{-1}$\,-\,416 s$^{-1}$, which was considered in a reasonable
agreement with the measured value $\Lambda_{1/2}\,=\,409\pm 40$
s$^{-1}$ \cite{CEAL}, in contrast to another measurement \cite{M}
providing somehow larger $\Lambda_{1/2}\,=\,470\pm 29$ s$^{-1}$. Let
us note that in these calculations the MECs are of the one-boson
exchange form. If used in conjunction with the nuclear wave
functions generated from one-boson exchange potentials (OBEPs), all
the parameters of the calculations are fixed. The same space
component of the weak axial MECs induces other very important weak
processes in the two-nucleon system, such as the weak neutrino- and
antineutrino-deuteron disintegration and the proton-proton fusion.
In the terrestrial conditions only the reaction (\ref{MUD}) has the
real hope to be studied experimentally with an accuracy of $\approx$
1 \% in the near future \cite{MSE}. This is challenge also for
theorists to improve the calculations of the capture rates. Let us
note that the above described approach was highly successful in
describing the electroweak processes in nuclei at low and
intermediate energies \cite{EW,CS}. We shall call it the Tree
Approximation Approach (TAA). It is also called the Standard Nuclear
Physics Approach \cite{APKM}.

A more fundamental approach to the study of electroweak processes in
nuclei was possible after the appearance of effective field theory
(EFT) \cite{W,GL}. In this way one constructs also terms of higher
order, including loops besides the leading order of a current
(potential) operator. The expansion is done in a perturbative
parameter $q/\Lambda_\chi\,<<\,1$, where $q$ is a quantity
characterizing the system (momentum, energy,...), which is small in
comparison with the heavy scale $\Lambda_\chi$.

Recently, the reaction (\ref{MUD}) was considered in two varieties
of such an EFT. In Ref.\cite{CIJL}, $\Lambda_{1/2}$ was calculated
within the framework of a pionless EFT. In this approach, an unknown
low energy constant (LEC) $L_{1,A}$ enters the weak axial MECs. It
cannot be extracted from the one-body processes nor derived from
fundamental considerations. In our opinion, the application of the
pionless EFT to the reaction (\ref{MUD}) is doubtful, because in
this process the contribution from the pion pole (the induced
pseudoscalar) is important. Since the pion is absent, Chen {\it et
al.} \cite{CIJL} introduced the pion pole by hands which is a
misconception. Moreover, the heavy scale turns out to be of the
order of the pion mass $m_\pi$, $\Lambda_\chi\,\sim\,m_\pi\,\sim$
\mbox{140 MeV}, whereas the value of the momentum transfer $q$ is of
the order of the muon mass $m_\mu$,  $q\,\sim\,m_\mu\,\sim$
\mbox{100 MeV}. The calculated prediction for $\Lambda_{1/2}$ for
reasonable values of $L_{1,A}\,\sim\,5-6$ (see FIG.\,3 and TABLE I
of Ref.\cite{CIJL}) is $\Lambda_{1/2}\,\sim\,370 - 380\,\rm{s}^{-1}$
MeV, which is by $\sim$ 5 - 10 \% less than the results obtained
within the above discussed TAA.

In Ref.\cite{APKM}, the muon capture in deuterium was studied within
a 'hybrid' approach: the current operator was constructed within the
framework of a heavy baryon EFT and the nuclear wave functions were
obtained from the Schroedinger equation solved with the second
generation Argonne $v_{18}$ potential \cite{WSS}. The weak axial
MECs operator of the heavy baryon EFT contains the one-pion exchange
part and also a (contact) short-range term, the strength of which is
given by the LEC $\hat d^R$, that should be fixed by data. In
Ref.\cite{APKM}, $\hat d^R$ is taken from the analysis of the triton
$\beta$-decay rate \cite{ASP} to predict the
$\Lambda_{1/2}(^{1}S_0)$ for the transition
$d\,\rightarrow\,^{1}S_0$. Let us note that the value of $\hat d^R$
is correlated with the cutoff $\Lambda$ entering a Gaussian
regulator (strong form factor). Since the Argonne potential is not a
OBEP containing such a form factor, the value of $\Lambda$ is not
fixed by the NN scattering data. Rather, the extraction of $\hat
d^R$ is made for three values $\Lambda\,$=\,500 MeV, 600 MeV and 800
MeV. For these values of $\hat d^R$ and $\Lambda$, the doublet
capture rate calculated in Ref.\,\cite{APKM} for the channel
$d\,\rightarrow\,^{1}S_0$, $\Lambda_{1/2}(^{1}S_0)\,=\,245\,$
s$^{-1}$, is by $\sim$ 5.5 \% less than the value
$\Lambda_{1/2}(^{1}S_0)\,=\,259\,$ s$^{-1}$, obtained in the TAA
calculations \cite{TKK}.

Here we calculate the capture rates for the reaction (\ref{MUD})
within both the TAA and the hybrid calculations. In both methods the
basic symmetries of quantum chromodynamics, such as the gauge
symmetry and the spontaneously broken chiral symmetry, are
reflected. The nuclear wave functions are obtained by solving the
Schroedinger equation \be H\Psi\,=\,E \Psi,\,\quad H\,=\,T+V\,,
\label{SE} \ee where $T$ is the the kinetic energy operator and $V$
is the NN potential. Since our TAA weak currents, constructed from
chiral invariant Lagrangians in Refs.\,\cite{IT1,IT2,ATCS,TK1,MRT1},
are of the OBE type, we use OBEPs in Eq.\,(\ref{SE}), too. Such
potentials are successfully applied in the TAA for describing
nuclear phenomena at low energies. This modeling of the nuclear
force is a surrogate of calculations based directly on the
quark-gluon dynamics because of the non-perturbative feature of the
quantum chromodynamics in this region of energies. The well known
first generation Bonn OBEPs \cite{MHE,MANUP} contain the exchanges
of scalar, pseudoscalar and vector bosons. In the BNN vertices,
phenomenological form factors of the form \be F_\alpha({\vec
q}^{\,\,2})\,=\,\left(\frac{\Lambda^2_\alpha\,-\,m_\alpha}
{\Lambda^2_\alpha\,+\,{\vec q}^{\,\,2}}\right)^{n_\alpha}
\label{PHFF} \ee are applied, with $\vec q$ the three-momentum
transfer, $\Lambda_\alpha$ the so called cutoff mass and
n$_\alpha$=1 or 2 depending on the specific couplings. These form
factors reflect the extended structure of the nucleon. Later on,
high quality second generation CD-Bonn OBEP appeared \cite{MCD},
containing the BNN form factors of the type (\ref{PHFF}), too.
Another type of OBEPs has been constructed by the Nijmegen group
\cite{SKTS}. These potentials contain the Gaussian strong form
factors, \be F_{BNN}(\vec q^{\,2})\,=\,e^{-\vec
q^{\,2}/2\Lambda^2_B}\,, \label{EFF} \ee where the cutoffs
$\Lambda_B$ are extracted from the fit to the NN scattering data.

In order to make our calculations of the capture rates consistent,
we employ in Eq.\,(\ref{SE}) the Nijmegen OBEPs and use the Gaussian
form factors (\ref{EFF}) and couplings entering these potentials
both in our weak and EFT MECs. As it has been shown in Section 3 of
Ref.\,\cite{ATA} the potential current of the range $B$,
$j^a_{\mu,\,B}(2,p.c.)$, of the weak vector nuclear MECs satisfies
the nuclear Conserved Vector Current (CVC) equation  \be q_\mu
j^a_{\mu,\,B}(2,p.c.)\,=\,[V_B,\,j^a_0(1)]\,, \label{NCVC} \ee where
$V_B$ is the OBEP of the same range $B$ and $j^a_0(1)$ is the
one-body vector charge density. In its turn, an analogous current of
the weak axial nuclear MECs, $j^a_{5\mu,\,B}(2,p.c.)$, fulfills (see
Section 3.1 and Appendix A of Ref.\,\cite{MRT1}) the nuclear
Partially Conserved Axial Current (PCAC) equation of the form  \be
q_\mu j^a_{5\mu,\,B}(2,p.c.)\,=\,[V_B,\,j^a_{50}(1)]\,+\,if_\pi
m^2_\pi\Delta^\pi_F(q^2){\cal M}^a_B(2)\,,  \label{NPCAC} \ee where
$j^a_{50}(1)$ is the one-body axial charge density and ${\cal
M}^a_B(2)$ is the associated two-nucleon pion absorption/production
amplitude. Further, $f_{\pi}$ is the pion decay constant and
$\Delta_F^{\pi}$ is the pion propagator. So the potential current in
this case depends only on the employed potential model. Our
definition of the MEC operators guarantees that there is no double
counting if the MEC effects are calculated with the wave functions
from Eq.\,(\ref{SE}).

Let us note that the presented approach has already been
successfully applied to the description of nuclear phenomena. For
example, in Refs.\,\cite{TS,STK}, the cross section of the reaction
of the backward deuteron disintegration, \be
e\,+\,d\,\rightarrow\,e'\,+\,n\,+\,p\,, \label{BED} \ee is well
described up to energies of 1 GeV \cite{S,GG}. In the weak sector,
the capture rate of the reaction \be
\mu^-\,+\,^{3}He\,\rightarrow\,\nu_\mu\,+\,^{3}H\,, \label{MCTHE}
\ee was calculated in Ref.\,\cite{CT}. The obtained value $\Gamma_0
= 1502 \pm 32$ s$^{-1}$ is in a very good agreement with the result
of the precise experiment \cite{AAV,PA}, $\Gamma^{exp}_0 = 1494.0
\pm 4.0 $ s$^{-1}$.

The procedure of implementing phenomenological strong form factors
into the electromagnetic MECs of the OBE type was in detail
discussed in conjunction with the Bethe-Salpeter equation in
Ref.\,\cite{GR}. Similar procedure for the weak axial MECs was
performed in Ref.\,\cite{KT}. The construction of our weak axial
nuclear MECs in conjunction with the Schroedinger equation
(\ref{SE}) stems in Ref.\,\cite{MRT1} from these currents. The set
of Eqs.\,(\ref{NCVC}) guarantees the gauge invariance of the
calculations with the weak vector current, whereas the set of
Eqs.\,(\ref{NPCAC}) provides the correct relation between the weak
axial current and the related pion production/absorption amplitude.
When one deals with potentials of particular form such as the OBEPs
containing phenomenological form factors, the splitting of the
continuity equation for the currents into the set of equations for
particular exchanges follows naturally. On the other hand, in the
given case and also for a potential V of general form one can
generate conserved currents in the weak vector sector using the
minimal substitution \cite{BLA,MVSKR}. For the weak axial sector, an
effective procedure for constructing the MECs satisfying the nuclear
form of the PCAC, \be q_\mu
j^a_{5\mu}(2,p.c.)\,=\,[V,\,j^a_{50}(1)]\,+\,if_\pi
m^2_\pi\Delta^\pi_F(q^2){\cal M}^a(2)\,,  \label{NPCACT} \ee is not
known.

Since the most important weak axial MECs are the $\Delta(1232)$
excitation currents of the $\pi$ and $\rho$ ranges, we pay
particular attention to them. At first we generalize them so that
also pieces depending explicitly on the off-shell parameters $Y$ and
$Z$ \cite{BDM} are included. Besides, we construct the $\Delta$
excitation currents also from the 'gauge symmetric' Lagrangians
\cite{PATI,PAS,SMT}. In calculations of the capture rates, we take
specific values of the parameters $Y$ and $Z$, which allows us to
study the model dependence. In particular, if we take $Y$ and $Z$ as
obtained from the requirement that the $\pi N \Delta$ and $\rho N
\Delta$ interactions should not change the number of degrees of
freedom of the free $\Delta$ \cite{LMN1,LMN2,PDT}, we obtain the
same capture rates as calculated with the $\Delta$ excitation
currents constructed from the gauge symmetric Lagrangians.

In our hybrid calculations, we apply the weak MECs that were used in
\cite{APKM}. However, we require that the weak vector MECs satisfy
Eq.\,(\ref{NCVC}) and that the leading order terms of the weak axial
MECs satisfy Eq.\,(\ref{NPCAC}), also for the Gaussian strong form
factor. In order that the weak axial MECs satisfy the nuclear form
of the PCAC (\ref{NPCAC}), one needs a potential current constructed
in \cite{MRT2}. In the weak vector sector, in addition to the pion
pair and pion-in-flight terms, we employ the $\Delta$ excitation
current of the $\pi$ and $\rho$ ranges. These vector MECs are well
known from the study of the process
$n\,+\,p\,\rightarrow\,d\,+\,\gamma$ at the threshold in the TAA
\cite{JFM}. Since the $\Delta$ excitation currents saturate about 30
\% of the MEC effect, we find it necessary to include them here,
too. Let us note that the strength of the $\Delta$ excitation
current of the $\pi$ range constructed within the TAA from the gauge
symmetric $\pi N \Delta$ Lagrangian is close to the strength of such
a current constructed within the heavy baryon EFT \cite{PMR}. The
inspection of Table 2 \cite{PMR} also confirms that this current
should contribute non-negligibly. As we shall see, our calculations
confirm that it is really the case.

In numerical calculations of the doublet capture rate, we use in the
hybrid approach the same value of the cutoff $\Lambda_\pi$ as in the
TAA, which allows us to extract from the comparison of the doublet
capture rates a unique value of the LEC $\hat d^R$.

In Section \ref{prel}, we discuss briefly the methods and inputs
necessary for the calculations, in Section \ref{cr} we display the
equations for the capture rates and in Section \ref{rd}, we present
the results. We conclude in Section \ref{concl}. Further, in
Appendix \ref{appA}, we present the weak currents of the TAA and
then we collect the Fourier transforms of the used MECs. In Appendix
\ref{appB} we consider analogously the EFT currents, and in Appendix
\ref{appC} we deliver all the multipoles of the currents used in the
numerical calculations of the capture rates.

\section{Methods and inputs}
\label{prel}

To obtain the capture rates one must calculate first of all the
matrix elements of the weak nuclear currents between the initial and
final nuclear states. Here we describe the needed ingredients of
these calculations.

\subsection{Weak nuclear currents of the TAA}
\label{wnctaa}

The weak hadron current, triggering the reaction (\ref{MUD}), is
\be j^-_{\,\mu}\,=\,j^-_\mu\,+\,j^-_{5\mu}\,. \label{WHC} \ee The
weak vector and weak axial nuclear currents $j^-_\mu$ and
$j^-_{5\mu}$, respectively,  consist of the one- and two-nucleon
parts presented in Appendix \ref{appA}. There is practically no
uncertainty associated with the one-body part. Hence we
concentrate on the effects of the two-body currents in the channel
$d\,\rightarrow\,^{1}S_0$.

The weak axial nuclear MEC $j^-_{5\mu}(2)$ that we consider here is
of the OBE-type with the $\pi$ and $\rho$  exchanges. It can be
divided \cite{MRT2} into the potential and non-potential currents.
The potential current of the range $B$, $j^-_{5\mu,B}(2,p.c.)$,
satisfies the nuclear PCAC equation (\ref{NPCAC}).

The main part of the non-potential weak axial exchange currents
contain the model independent $\rho$-$\pi$ current and the $\Delta$
excitation currents that are model dependent. In our calculations,
we shall adopt the $\pi N \Delta$ and $\rho N \Delta$ Lagrangians
used for many years \cite{BDM,OO} to study the $\pi N$ reactions and
the pion photo- and electroproduction on a nucleon (model I) and
also the gauge symmetric Lagrangians proposed recently
\cite{PATI,PAS}. In model I, we derived the $\Delta$ excitation MECs
from Lagrangians possessing the hidden local $SU(2)_L\times SU(2)_R$
symmetry \cite{SMT}. In particular, the vertices containing the
$\Delta$ isobar field were chosen as \be {\cal L}_{\,N\Delta\pi\rho
a_{1}} \, =\,{\cal L}_{\,N\Delta\pi a_1}\,+\,{\cal
L}^1_{\,N\Delta\rho}\,+\,{\cal L}^2_{\,N \Delta\rho}\,,
\label{LNDPRA1} \ee where \bea {\cal L}_{\,N\Delta\pi a_{1}} \,&
=&\,\frac{f_{\pi N \Delta}} {m_{\pi}}\,\bar{\Psi}_{\mu}\vec{T}{\cal
O}_{\mu\nu}(Z)\Psi\cdot
\left(\partial_{\nu}\vec{\pi}+2 f_{\pi}g_\rho\vec{a}_{\nu}\right)\,+\,h.\,c.\,, \label{LPIND} \\
{\cal L}^1_{\,N\Delta\rho}\,&=&\,
-g_{\rho}\frac{G_{1}}{M}\bar{\Psi}_{\mu}\vec{T}{\cal
O}_{\mu\eta}(Y) \gamma_{5}\gamma_{\nu}\Psi\cdot
\vec{\rho}_{\eta\nu}\,+\,h.\,c.\,, \label{L1RND} \\
{\cal L}^2_{\,N\Delta\rho}\,&=&\,
g_{\rho}\frac{G_{2}}{M^2}\bar{\Psi}_{\mu}\vec{T}{\cal
O}_{\mu\eta}(X) (\partial_{\nu}\Psi)\cdot
\vec{\rho}_{\eta\nu}\,+\,h.\,c.\,. \label{L2RND} \eea Here $\vec{T}$
is the operator of the isospin $1/2\,\rightarrow\,3/2$ transition.
The constant $f_{\pi N \Delta}$ is extracted from the $\Delta$
isobar width \cite{PDG} with the result \be \left(\frac{f_{\pi N
\Delta}} {m_{\pi}}\right)^2/4\pi=0.767\,\pm\,0.024\, \rm{fm}^2\,.
\label{FPIND} \ee Besides, $G_1=2.525$ \cite{BDM}. Further, the
operator ${\cal O}_{\mu\nu}(B)$ is taken in the form
\cite{BDM,OO,DMW1,DMW2} \bea
{\cal O}_{\mu\nu}(B)\,&=&\,\delta_{\mu\nu}\,+\,C(B)\,\gamma_\mu\,\gamma_\nu\,, \label{OmnB} \\
C(B)\,&=&\,-\left(\frac{1}{2}\,+\,B\right)\,.  \label{CeB} \eea
The parameters $X$, $Y$ and $Z$ do not influence the on--shell
properties of the $\Delta$ isobar, hence they are called off-shell
parameters. They were systematically extracted from the data
in Refs.\,\cite{BDM,OO,DMW1,DMW2}.

On the other hand, the parameters entering the Lagrangians
(\ref{LNDPRA1}) were restricted in \cite{LMN1,LMN2} on the basis of
field theoretical arguments by the requirement that they should
conserve the same number of degrees of freedom of the $\Delta$
isobar as possessed by the free one. This resulted in that it should
be \be Y=0\,,\quad Z=\fot\,\,, \quad G_2=0\,. \label{PARRES} \ee
This was criticized in \cite{BDM}, because the requirement $G_2=0$
fixes the ratio of the multipole E2 to multipole M1 amplitudes
kinematically, which is unacceptable and the restriction
(\ref{PARRES}) was refused as a whole. In our opinion the
consistency requirement $G_2=0$ means that the model I can be only
applied if the amplitudes derived from the Lagrangian ${\cal
L}^2_{\,N\Delta\rho}$, Eq.\,(\ref{L2RND}), do not contribute.

Later on, the problems related to the Lagrangians (\ref{LNDPRA1})
were reconsidered in \cite{PDT,PATI,PAS}. In particular it was shown
in \cite{PDT} within the framework of the Hamiltonian formalism with
constraints that the result $Z=\fot$ persists.

Since the contribution to the effects of the $\Delta$ excitation
currents for the process (\ref{MUD}) from the Lagrangian
(\ref{L2RND}) is negligible, we consider in our calculations two
sets of parameters $Y$ and $Z$. In the set Ia we take \be
Y\,=\,Z\,=\,-\fot\,.  \label{SIa} \ee In its turn, the set Ib is
defined as \be Y\,=\,0\,, \quad Z\,=\,\fot\,.  \label{SIb} \ee The
$\Delta$ excitation currents corresponding to the set (\ref{SIa})
were used in all previous calculations of the MECs effects for the
reaction (\ref{MUD}).

The $\pi N \Delta$ and $\rho N \Delta$ Lagrangians of the model II
\cite{PATI,PAS} do not contain any off-shell parameters. They are
constructed in such a way that they conserve the degrees of freedom
of the free $\Delta$ isobar. We write them as \cite{SMT} \be {\cal
L}^{g.s.}_{\,N\Delta\pi a_1} \, = \,\frac{f_{\pi N \Delta}}{m_\pi
M_\Delta}\,\vep_{\mu\nu\alpha\beta}
\,[(\partial_\mu\bar{\Psi}_{\nu})\,\vec{T}\gamma_5\gamma_\alpha\,\Psi]\cdot
(\partial_{\beta}\vec{\pi}\,+\,2f_\pi g_\rho {\vec
a}_\beta)\,+\,h.c.\,,   \label{LGSPND} \ee \be {\cal
L}^{g.s.}_{\,N\Delta\rho } \, =
\,\frac{G_1}{MM_\Delta}\,g_\rho\,\left\{\,\vep_{\mu\nu\alpha\beta}
\,[(\partial_\mu\bar{\Psi}_{\nu})\,\vec{T}\gamma_\alpha\gamma_\lambda\,\Psi]\,
+\,[(\partial_\mu\bar{\Psi}_{\beta}-\partial_\beta\bar{\Psi}_{\mu}){\vec
T}\gamma_5\gamma_\mu\gamma_\lambda\,\Psi]\right\} \cdot{\vec
\rho}_{\lambda\beta}\,+\,h.c.\,.  \label{LGSRND} \ee The values of
the coupling constants are obtained from the condition that the new
Lagrangians, Eq.\,(\ref{LGSPND}) and Eq.\,(\ref{LGSRND}) and the
standard  Lagrangians, Eq.\,(\ref{LPIND}) and Eq.\,(\ref{L1RND}),
respectively, are equivalent for the on--shell $\Delta$ isobar.

It turns out that the $\Delta$ excitation currents of this model
differ from those of the model I, set Ia (\ref{SIa}), only by the
factor $(M/M_\Delta)^2\,\approx\,0.58$ [$M (M_\Delta)$ is the
nucleon ($\Delta$ isobar) mass].

\subsection{Weak nuclear currents of the EFT approach}
\label{wnceft}

For the estimation of the MECs effect in the hybrid calculations, we
use the MECs operator of Ref.\,\cite{APKM} from which we omit the
second term at the right hand side of Eq.\,(19). This term is
suppressed by the factor $1/M$ in comparison with the leading term,
which provides by itself only a small contribution to the weak axial
MECs effect. In the weak vector sector, Ando {\it et al.}
\cite{APKM} present in Eq.\,(17) the non-relativistic version of the
currents $J^\mu_{em}(a)$ and $J^\mu_{em}(b)$ of the set (C.4),
constructed within the heavy baryon EFT \cite{PMR}. They coincide
with our $\pi$-pair term (\ref{VPPT}) and pion-in-flight term
(\ref{VPFT}), constructed within the formalism of hard pion
Lagrangians. Here we add to them the $\Delta$ excitation currents of
the $\pi$ and $\rho$ ranges (\ref{VDCPIR}) and (\ref{VDCRHOR}). It
can be shown that the $\Delta$ excitation current of the $\pi$ range
(\ref{VDCPIR}), if taken with the set (\ref{SIa}) and multiplied by
the factor $(M/M_\Delta)^2$ corresponds to the $\Delta$ excitation
current $J^\mu_{em}(f)$, Eq.\,(C.4) of Ref.\,\cite{PMR}. The TAA
calculations show that the currents (\ref{VDCPIR}) and
(\ref{VDCRHOR}) contribute non-negligibly. So in order to assess the
capture rate $\Lambda_{1/2}(d\,\rightarrow\,^{1}S_0)$ correctly, one
should include explicitly also these currents in the EFT calculation
of the capture rates. In our opinion the current (\ref{VDCRHOR}) can
be obtained from the heavy-fermion Lagrangians of Appendix C
\cite{PMR} as well.

In the weak axial sector, we add to the weak axial MECs \cite{APKM}
the $\pi$ potential term (\ref{APIPT}). This ensures that also the
weak axial MECs of the EFT approach satisfy in the leading order the
PCAC constraint (\ref{NPCAC}) \cite{MRT2}.

The weak axial MECs of the heavy baryon EFT contain the known LECs
$\hat c_i,$ ($i=1,2,3,4$), $c_6$ and the unknown LECs $\hat d_1$ and
$\hat d_2$. These two constants enter in the calculations of the
observables effectively in the combination $\hat d_1+2\hat d_2$ or
as used in the formalism of Ref.\,\cite{APKM} in the combination \be
\hat d^R\,\equiv\,\hat d_1+2\hat d_2+\frac{1}{3}\hat
c_3+\frac{2}{3}\hat c_4+\frac{1}{6}\,. \label{HDR} \ee On the other
hand the combination $\hat d_1+2\hat d_2$ is also expressed in terms
of the LEC $c_D$ as \cite{GQN} \be \hat d_1+2\hat
d_2=-\frac{M_N}{\Lambda_\chi g_A} c_D\,, \label{CED} \ee where
M$_N$=0.939 GeV is the nucleon mass and $\Lambda_\chi \approx$ 1
GeV.

Let us note that the constant $\hat d^R$ (c$_D$) enters not only the
short range part of the weak axial MEC but also the pion
production/absorption amplitude in the NN collisions and the NNN
force.

The dependence of the LECs on the off-shell parameter $Z$ was
discussed by Bernard {\it et al.} \cite{BKM1} in the model where the
LECs are saturated by the heavy mesons and the $\Delta$ resonance.
In parallel with the model Ia (\ref{SIa}), in which $Z=-\fot$ we
obtain the set IIa \be \hat c_2=3.37\,,\quad \hat c_3=-4.70\,,\quad
\hat c_4=3.31\,,\quad\rm{(IIa)}\,, \label{RSIa} \ee whereas for the
case Ib (\ref{SIb}) in which $Z=\fot$ the LECs of the model IIb are
\be \hat c_2=1.98\,,\quad \hat c_3=-3.31\,,\quad \hat
c_4=2.61\,,\quad\rm{(IIb)}\,. \label{RSIb} \ee

Another set of these constants, which we design as the set IIc, has
been extracted in \cite{BKM2} from the data, \be \hat c_2=1.67\pm
0.09\,,\quad \hat c_3=-3.66\pm 0.08\,,\quad \hat c_4=2.11\pm
0.08\,,\quad\rm{(IIc)}\,. \label{CIEX} \ee  If one omits from the
LECs $\hat c_2$, $\hat c_3$ and $\hat c_4$ the contribution from the
$\Delta$ isobar, one gets the model IId \be \hat c_2=0.047\,, \quad
\hat c_3=-1.371\,,\quad \hat c_4=1.63\,,\quad\rm{(IId)}\,.
\label{CWD} \ee For the constants $\hat c_1$ and $c_6$ we take
\cite{BKM2,APKM} \be \hat c_1=-0.60\pm 0.13\,,\quad  c_6=3.70\,.
\label{C16} \ee New set of the LECs has recently been delivered in
Ref.\,\cite{VB}. We take it as model IIe, \be \hat
c_1=-0.85\,\begin{array}{c}
                         +0.2 \\ -0.5 \end{array}\,, \quad \hat c_2=3.1\pm 0.2\,, \quad \hat c_3=-4.4
\begin{array}{c}
                         +1.2 \\ -1.0 \end{array}\,,\quad \hat c_4=3.3 \begin{array}{c}
                         +0.5 \\ -0.2 \end{array}\,,\quad\rm{(IIe)}\,. \label{CIEXVB} \ee
These new LECs are close to the set IIa (\ref{RSIa}) obtained for $Z=-\fot$.
However, these new LECs  are extracted with much larger errors than
the set IIc (\ref{CIEX}).

\subsection{Nuclear potentials}
\label{np}

We use the Nijmegen I (NI) and Nijmegen 93 (N93) \cite{SKTS} OBEPs.
The couplings and cutoffs, entering these potentials, are employed
also in the MECs. In particular, the cutoffs are \be \Lambda_\pi=
827.5\,\rm{(NI)}\,, \quad \Lambda_\pi=1177.11\,\rm{(N93)}\,.
\label{LAMC} \ee Let us note that our main results are related to
the high quality second generation NI potential with
$\chi^2$/N$_{data}$=1.03, whereas for the N93 potential
$\chi^2$/N$_{data}$=1.87.

\section{Capture rates}
\label{cr}
The capture rates $\Lambda_{1/2}$ and $\Lambda_{3/2}$ are related to
the statistical capture rate,
$\Lambda_{stat}$, as
\be \Lambda_{stat}\,=\,\frac{1}{3}\,\Lambda_{1/2}\,+\,\frac{2}{3}\,\Lambda_{3/2}\,. \label{LSTAT0}
\ee
Using the method of Ref.\,\cite{WAL}, we write
the statistical capture rate for the process (\ref{MUD}) in terms of the multipoles,
\bea \Lambda_{stat}\,&=&\,\frac{M_n}{3}\,\bigg[\frac{G_F \cos \theta_C \phi_\mu(0)}{\pi}\bigg]^2\,
\int^{\nu_{max}}_0\,d\nu\,(\nu^2/\kappa_0) \,\sum_{\lambda j_f,\,J}
\left\{\,|<\lambda j_f||i\hat T^{\,el}_J-\hat T^{\,mag}_J||d>|^2  \right. \nonumber \\
&& \left. \,+\,|<\lambda j_f||{\hat L}_J-{\hat M}_J||d>|^2\,
\right\}\,,\label{LSTAT1} \eea where \bea \kappa_0 &=&
\sqrt{M_n(\Delta-\nu-\nu^2/4M_n)}\,, \quad
\nu_{max}=2M_n(-1+\sqrt{1+\Delta/M_n})\,, \nonumber \\
\Delta &=& m_\mu+m_d-2M_n-|\epsilon_\mu|\,, \label{VAR} \eea
$M_n\,(m_d)$ is the neutron (deuteron) mass, and
$\epsilon_\mu=-0.00267$ MeV is the binding energy of the muon.
Further in (\ref{LSTAT1}), the weak interaction constant
$G_F=1.16637\times 10^{-5}$ GeV$^{-2}$ \cite{PDG}, $\cos
\theta_C$=0.9749 and the wave function of the bound muon at the
center of the deuteron is
$\phi_\mu(0)=(m_{\mu\,,r}\,\alpha)^{3/2}/\sqrt{\pi},$ $m_{\mu\,,r}$
is the reduced mass of the muon and $\alpha$ is the fine structure
constant.

The multipoles and their reduced matrix elements are defined in Section 4.1 of Ref.\,\cite{MRT1}.

Generally, the capture rates $\Lambda_F$ for the hyperfine state $F$
can be obtained from the equation \cite{WAL}, \be
\Lambda_F\,=\,\Lambda_{stat}\,+\,C_F\,\delta\Lambda\,, \label{LF}
\ee where \be
C_F\,=\,(-1)^{F+\fot}\,\seij{J_i}{\fot}{F}{\fot}{J_i}{1}\,,
\label{CF} \ee and $\delta\Lambda$ depends on the nuclear dynamics.
Here J$_i$ is the total angular momentum of the initial nucleus.

For the reaction (\ref{MUD}), J$_i$=1 and \be
C_\fot\,=\,\frac{1}{3}\,,\quad C_{\frac{3}{2}}\,=\,-\frac{1}{6}\,,
\label{CS} \ee whereas \bea
\delta\Lambda\,&=&\,\sqrt{6}M_n\,\bigg[\frac{G_F \cos \theta_C
\phi_\mu(0)}{\pi}\bigg]^2\,
\int^{\nu_{max}}_0\,d\nu\,(\nu^2/\kappa_0)\,\sum_{\lambda
j_f,\,J\,J'}\,
{\hat J}\,{\hat J'}\,(-1)^{j_f}\seij{J_i}{J}{j_f}{J'}{J_i}{1} \nonumber \\
&&\,\left\{\,i^{J-J'}\,\trej{J}{J'}{1}{-1}{1}{0}<\lambda j_f||i\hat T^{\,el}_J-\hat T^{\,mag}_J||d>
<\lambda j_f||i\hat T^{\,el}_{J'}-\hat T^{\,mag}_{J'}||d>^* \right. \nonumber \\
&&\left.\,-2\sqrt{2}\trej{J}{J'}{1}{-1}{0}{1}\,\Re\bigg[\,i^{J-J'}\,<\lambda j_f||i\hat T^{\,el}_J-\hat T^{\,mag}_J||d>
<\lambda j_f||{\hat L}_{J'}-{\hat M}_{J'}||d>^*\,\bigg] \right. \nonumber \\
&&\left.\,+\,i^{J-J'}\,\trejzero{J}{J'}{1}\,<\lambda j_f||{\hat
L}_J-{\hat M}_J||d> <\lambda j_f||{\hat L}_{J'}-{\hat
M}_{J'}||d>^*\,\right\}\,. \label{DL} \eea Here $\hat J=\sqrt{2J+1}$
and the symbols $\trej{j_1}{j_2}{j_3}{m_1}{m_2}{m_3}$ and
$\seij{j_1}{j_2}{j_3}{j_4}{j_5}{j_6}$ are Wigner's  $3jm$ and $6j$
symbols, respectively \cite{VMK}.

\section{Results and discussion}
\label{rd}

Here we present first the results for the capture rates obtained for
the reaction (\ref{MUD}) in the formalism of the TAA and then in the
hybrid calculations.

\subsection{Results for the TAA}
\label{taac}
In the formalism of the TAA, we calculated the contributions to the capture rates from all
channels $d\,\rightarrow\,^{2S+1}L_{j_f},$ where L=S,P,D,F, j$_f$=0,1,2 and
from the multipoles J=0,1,2,3 of the one-nucleon currents. The contribution of the weak MECs
was taken into account in the multipole J=1 and in the channel $d\,\rightarrow\,^{1}S_0$.
We also estimated the MEC effect due to the weak vector MECs $\vec j^{\,\,-}(p.t.)$, Eq.\,(\ref{VPPT}), and
$\vec j^{\,\,-}(\pi\pi)$, Eq.\,(\ref{VPFT}) in the channels $d\,\rightarrow\,^{3}P_{0,1,2}$.

The results of the calculations of the doublet capture rate
$\Lambda_{1/2}$ for the reaction (\ref{MUD}) are presented in Table
\ref{tab:dtr}.

\begin{table}[htb]
\caption{Partial contributions to $\Lambda_{1/2}$ (in s$^{-1}$). The
value of the constant $\left(\frac{f_{\pi N
\Delta}}{m_{\pi}}\right)^2/4\pi=0.783$ fm$^2$ is used. In the first
column, the potential and the current model are displayed. In the
second column, the contribution from the one-body current in the
channel $d\,\rightarrow\,^{1}S_0$ is given. In the third column, the
contribution of the MECs in the channel $d\,\rightarrow\,^{1}S_0$ is
presented, whereas in the fourth column, the contribution of the
one-body current to all considered channels but
$d\,\rightarrow\,^{1}S_0$ is given. In the fifth column, we give the
total MECs effect and in the last column, all the contributions are
summed up.}
\begin{center}
\begin{tabular}{|l | c   c  c  c  c  |}\hline
         & IA$_0$ & $\Delta$MEC$_{0}$ & $\Delta$IA &  MEC & IA+MEC \\\hline
NI/Ia    & 239.2  & 22.0 & 160.4 & 23.8 & 423.4 \\
NI/Ib    & 239.2  & 14.8 & 160.4 & 16.7 & 416.3 \\
N93/Ia   & 238.8  & 29.1 & 160.4 & 30.8 & 430.0 \\
N93/Ib   & 238.8  & 22.0 & 160.4 & 23.8 & 423.0 \\\hline
\end{tabular}
\end{center}
\label{tab:dtr}
\end{table}
It is seen from Table \ref{tab:dtr} that the results depend on the
potential and current model used. As noted above, our basic model is
NI/Ib: the potential NI qualitatively supersedes the N93 one and the
current model Ib should be preferred because the parameters of the
$\Delta$ excitation currents are restricted by the reasonable demand
that the number of degrees of freedom of the $\Delta$ isobar should
be conserved. Moreover, the same results as displayed in the third
row of Table \ref{tab:dtr} are obtained with the potential NI and
with the current model based on the Lagrangians (\ref{LPIND}) and
(\ref{L1RND}). As is seen from the fifth column of the Table
\ref{tab:dtr}, the resulting MECs effect for this case is 16.7
s$^{-1}$, which is \mbox{$\approx$ 4 \%}. The estimated error in
$\Lambda_{1/2}$ for this model due to the 3 \% variation of the
constant $\left(\frac{f_{\pi N \Delta}}{m_{\pi}}\right)^2/4\pi$
given in Eq.\,(\ref{FPIND}) is $\approx$ 1 s$^{-1}$, which is
$\approx$ 0.25 \%.

Let us note that the time component of the weak axial MECs
contributes by +2.3 (+2.9) s$^{-1}$ for the models NI/Ia and NI/Ib
(N93/Ia and N93/Ib). Our calculations show that the effect of the
weak vector $\Delta$ excitation currents is $\approx$ 4 -- 6
s$^{-1}$. It is a non-negligible contribution to $\Lambda_{1/2}$ of
the order of 1 -- 1.5 \%, which cannot be neglected if the expected
error of the data is $\approx$ 1.5 \%. If one looks for the effect
of the heavy meson exchanges, one obtains for the model NI/Ib that
\be \Delta
\Lambda^{\rm{MEC}}_{1/2}\,=\,24.3\,\rm{s}^{-1}\,-\,7.6\,\rm{s}^{-1}\,=\,16.7
\,\rm{s}^{-1}\,, \label{HME} \ee where the first number at the
right-hand side of the equation is due to the MECs of the pion range
and the second number stems from the heavy meson exchanges. So the
effect of the short range MECs is $\approx$ 30 \% of the long range
one.

In Table \ref{tab:ia} we present the contribution from particular
multipoles to the capture rates calculated with the IA currents. It
is seen that the contributions from higher multipoles not taken into
account cannot change the results much.
\begin{table}[htb]
\caption{Partial contributions to the capture rates (in s$^{-1}$)
from the multipoles J, calculated in the IA approximation. The
neutron-neutron $^{2S+1}L_{j_f}$ partial waves with L=0,1,2,3 and
j$_f$=0,1,2 are taken into account. In the second column, only the
contribution from the channel $d\rightarrow ^{1}S_0$ to the
multipole J=1 is considered. In the third column, the capture rates
are calculated with the multipoles J=0 added; in the fourth column,
all the multipoles with J=1 but arising from the channel
$d\rightarrow ^{1}S_0$ are added; in the fifth (sixth) column, the
multipoles J=2 (J=3) are added.}
\begin{center}
\begin{tabular}{|l | c  c  c  c  c  |}\hline
  J              & 1 ($^{1}S_0$)  &    0     &    1    &    2    &    3     \\\hline
$\Lambda_{stat}$ &     83.2       &   93.2   & 119.6   & 139.4   &   141.0  \\
$\Lambda_{1/2}$  &    239.2       &  249.3   & 322.4   & 394.6   &   399.6  \\
$\Lambda_{3/2}$  &      5.1       &   15.2   &  18.3   &  11.9   &    11.7  \\\hline
\end{tabular}
\end{center}
\label{tab:ia}
\end{table}

We now give for the model NI/Ib the final results for the capture rates. In the channel
$d\,\rightarrow\,^{1}S_0$,
\be \Lambda^0_{stat}\,=\,88.1\,\rm{s}^{-1}\,,\quad \Lambda^0_{1/2}\,=\,254.0\,\rm{s}^{-1}\,,\quad
\Lambda^0_{3/2}\,=\,5.2\,\rm{s}^{-1}\,. \label{OSZLS} \ee

The full calculations provide
\be \Lambda_{stat}\,=\,146.4\,\rm{s}^{-1}\,,\quad \Lambda_{1/2}\,=\,416.3\,\rm{s}^{-1}\,,\quad
\Lambda_{3/2}\,=\,11.4\,\rm{s}^{-1}\,. \label{TLS} \ee

The result $\Lambda_{1/2}\,=\,416.3\,\rm{s}^{-1}$, Eq.\,(\ref{TLS}),
seems to be in agreement with the one of Eq.\,(38) of
Ref.\,\cite{ATCS}, $\Lambda_{1/2}= 416\pm 7\,\rm{s}^{-1}$. However,
it is more correct to compare this result with the value
$\Lambda_{1/2}$ = 430 $\rm{s}^{-1}$ of our Table \ref{tab:dtr},
obtained from the first generation realistic potential N93 and the
TAA current model Ia, because this sort of potentials and of the
current model was used also in \cite{ATCS}. In Ref.\,\cite{TKK}, the
value $\Lambda_{1/2} \approx 400\,\rm{s}^{-1}$ was reported, as the
result of calculations with similar potentials and currents as in
\cite{ATCS}. On the other hand, according to Ref.\,\cite{APKM}, the
contribution of the higher partial waves was later reevaluated
\cite{TKK} and an enhancement of the $\Lambda_{1/2}$ by $\approx$ 10
$\rm{s}^{-1}$ was achieved. Then  $\Lambda_{1/2} \,\approx$ 410
$\rm{s}^{-1}$ is in good agreement with the result of \cite{ATCS},
but it is by 5 \% smaller than our corresponding value of
$\Lambda_{1/2}= 430 \rm{s}^{-1}$. Let us note the calculations of
Ref.\,\cite{DSOM} reporting $\Lambda_{1/2}\,=\,402\,\rm{s}^{-1}$,
also performed with the first generation realistic potentials and
the current model Ia.

\subsection{Results for the EFT currents}
\label{eftc}

Here we provide the results of calculations of the LECs $\hat d^R$
and c$_D$ by comparing the doublet transition rate $\Lambda^0_{1/2}$
calculated with the weak MECs of Section \ref{taac}. We made the
calculations with the potential NI and the weak MECs discussed in
Section \ref{wnceft}. Using the Goldberger-Treiman relation we
connected the coupling of MECs \cite{APKM} with the constant g$_{\pi
NN}$ that we took from the potential. Since the cutoff $\Lambda_\pi$
is also taken from the potential, our hybrid calculations are
consistent as much as possible. As in Ref.\,\cite{APKM}, we use here
the weak form factors in the linear approximation in the expansion
in the four momentum transfer $q^2$, given in Eqs.\,(\ref{FV1QRA}),
and (\ref{FAQRA}). However, this provides only a small difference in
the results, in comparison with the full $q^2$ dependence.

It follows from Table \ref{tab:dtr} for the model NI/Ia that
$\Lambda^0_{1/2}=261.2$ s$^{-1}$. This provides for the model NI/IIa
the value of the LEC $\hat d^R=3.225$. In both models, $Z=-\fot$.
For the other considered models, we take $\Lambda^0_{1/2}=254.0$
s$^{-1}$ obtained for the model NI/Ib and the resulting constants
$\hat d^R$ and c$_D$ are presented in Table \ref{tab:hdr}.

\begin{table}[htb]
\caption{Values of the LECs $\hat d^R$ and $c_D$ obtained by
comparing the doublet capture rate for the channel
$d\,\rightarrow\,^{1}S_0$, calculated in Section \ref{taac} for the
model NI/Ib, and the one calculated with the weak MECs of Section
\ref{wnceft}. The value $Z=\fot$ is taken in the model NI/Ib and
also in the model NI/IIb, in calculating the contribution of the
$\Delta$ resonance to the LECs $\hat c_i,\,i=2,3,4$. In the model
NI/IId, the contribution of the $\Delta$ resonance to the LECs $\hat
c_i,\,i=2,3,4$ is omitted, whereas the $\Delta$ excitation current
of the pion range is explicitly taken into account.}
\begin{center}
\begin{tabular}{|    l     |     c      c      c    c   |}\hline
                           & NI/IIb  & NI/IIc & NI/IId & NI/IIe \\\hline
                $\hat d^R$ &  2.410  &  2.155 &  0.625 &  2.680 \\
                   c$_D$   &  2.173  &  2.436 & -0.231 &  2.407 \\\hline
\end{tabular}
\end{center}
\label{tab:hdr}
\end{table}
Comparing the second and the fourth columns of Table \ref{tab:hdr}
shows that taking into account the $\Delta$ resonance effect by the
method of the resonance saturation of the LECs and taking it into
account explicitly by calculating the MECs effect is not equivalent.
Also comparing the third and the last columns one finds about
\mbox{20 \%} change in the value of the constant $\hat d^R$. Having
in mind the large uncertainty in the LECs of the set IIe
(\ref{CIEXVB}) one concludes that the change is not essential. Let
us also note that the value $\hat d^R$=2.68 was obtained with
$Y=Z=-\fot$ in the $\Delta$ excitation currents. If one employs
$Y=0$ and $Z=-\fot$, the value of $\hat d^R$=2.66, so it changes
insignificantly. The value of the $\Lambda_\pi$ entering the NI
potential (\ref{LAMC}) is close to the value of one of the cutoffs,
$\Lambda$=800 MeV, used in the analysis of the triton beta decay
\cite{ASP}. For this value of the cutoff, the extracted  \mbox{$\hat
d^R=3.90\pm 0.10$} \cite{ASP}, which is enhanced at least by 30 \%
in comparison with $\hat d^R$ from our Table \ref{tab:hdr}.

We have also calculated the influence of the weak axial potential
current ${\vec j}^a_{5\mu,\pi}(2,p.c.)$ on the value of the constant
$\hat d^R$. As discussed in Ref.\,\cite{MRT2}, this current is
usually absent in calculations of the weak processes. If we omit it
the value of the constant $\hat d^R$=2.410 in the second column of
Table \ref{tab:hdr} increases to $\hat d^R$=2.740, thus it changes
by \mbox{$\approx$ 13 \%}. It follows that if one would like to
extract the value of $\hat d^R$ with an accuracy better than 10 \%,
then one should take the contribution of the potential current into
account. The calculations also show that omitting the potential
current causes an enhancement of the doublet transition rate
$\Lambda_{1/2}$ by \mbox{$\approx$ 1 \%}.

The constant c$_D$ has recently been extracted \cite{GQN}, together
with another LEC c$_E$, entering the contact part of the NNN force,
from the data on the triton beta decay, binding energies and
point-proton radii of the 3N system and $^{4}$He nucleus, with
resulting value \mbox{c$_D$=-0.2.} As it is seen from our Table
\ref{tab:hdr}, only the value \mbox{c$_D$=-0.231}, corresponding to
the model NI/IId, is in agreement with the analysis of
Ref.\,\cite{GQN}. Let us note that with the choice \mbox{$\hat
c_4=-\hat c_3$=3.4} \cite{GQN} one obtains  the value \mbox{$\hat
d^R$=1.1} using \mbox{c$_D$=-0.2.}

It follows from our calculations that the effect of the time
component of the weak axial MECs is $\approx$ -1 s$^{-1}$, which is
in agreement with Ref.\,\cite{APKM}. This is in contrast to the TAA
calculations based on the hard pion Lagrangians, where the time
component contributes as $\approx$ +2 s$^{-1}$. The short range part
of the hard pion time component reduces the value $\hat d^R$= 2.155
(see the third column of Table \ref{tab:hdr}) to $\hat d^R$= 1.91,
if one uses this component instead of the soft pion one in fitting
$\hat d^R$.

Our $\Lambda^0_{1/2}=254.0$ s$^{-1}$ differs by  9 s$^{-1}$ from the
same quantity, $\Gamma^{L=0}_{\mu d}=245$ s$^{-1}$, given in Table 1
of Ref.\,\cite{APKM}. The main part of the difference can be
assigned to the difference of $\approx$ 7 s$^{-1}$ in the
contribution from the one-body currents, obtained from comparing
IA$_0$=239.2 s$^{-1}$ (see our Table \ref{tab:dtr}) with the value
of $\Gamma_{\mu d}$=232 s$^{-1}$ of Table 2 \cite{APKM}. On the
other hand, our MECs effects, $\Delta$MEC$_{0}$=14.8 s$^{-1}$, are
close to 13 s$^{-1}$ obtained in \cite{APKM}.

\section{Conclusions}
\label{concl}

We have evaluated the capture rates for the reaction of muon capture
in deuterium (\ref{MUD}), both using the TAA currents and those
derived within the EFT approach.

The weak TAA currents, presented in Appendix \ref{appA}, are of the
one-boson exchange type obtained from the hard pion chiral
Lagrangians and they satisfy the nuclear CVC and PCAC constraints,
Eqs.\,(\ref{NCVC}) and (\ref{NPCAC}), respectively. The final state
neutron-neutron wave functions were generated from the high quality
second generation potential NI and from the realistic potential N93
\cite{SKTS}. Since the potentials are also of the one-boson exchange
type, employing in the TAA currents the same couplings and strong
form factors (\ref{EFF}), we performed fully consistent
calculations, presented in Section \ref{taac}. For the main object
of interest, the doublet capture rate $\Lambda^0_{1/2}$ for the
channel $d\rightarrow ^{1}S_0$, we predict (see Table \ref{tab:dtr},
model NI/Ib) \be \Lambda^0_{1/2}\,=254\,\pm\,3\,\rm{s}^{-1}\,.
\label{LOHF1S0} \ee This result was obtained with the
neutron-neutron wave functions derived from the NI potential. The
error reflects the uncertainty in the $\pi N \Delta$ and $\rho N
\Delta$ couplings, possible effects of the neglected short range
effects and applied approximations. In the model NI/Ib, the $\pi N
\Delta$ and $\rho N \Delta$ couplings preserve the physical degrees
of freedom of the free $\Delta$ isobar. The IA currents contribute
to the value (\ref{LOHF1S0}) of $\Lambda^0_{1/2}$ by 239 s$^{-1}$,
whereas the MEC effect is 15 s$^{-1}$, which is $\approx$ 6 \%. In
the full calculations, we considered the neutron-neutron
$^{2S+1}L_{j_f}$ partial waves with L=0,1,2,3 and j$_f$=0,1,2, and
the contributions of the IA currents to the multipoles J=0,1,2,3.
For the full doublet capture rate we got \be
\Lambda_{1/2}\,=416\,\pm\,6\,\rm{s}^{-1}\,.  \label{LOHF} \ee In
addition, the estimated error includes also the uncertainty due to
the neglect of the contribution from the higher multipoles.

The EFT currents that we used are discussed in Appendix \ref{appB}.
The hybrid calculations of the capture rates accomplished with these
currents are presented in Section \ref{eftc}. These calculations are
consistent to the extent that we again use in the MECs the couplings
and strong form factors from the potential NI. We extract the
unknown LEC ${\hat d}^R$ by comparing the capture rate
$\Lambda^0_{1/2}$ with its numerical value calculated with the TAA
currents. Besides, we adopt various available sets of the known LECs
${\hat c}_i$, discussed in Section \ref{wnceft}. As is seen from
Table \ref{tab:hdr}, the value of ${\hat d}^R$ changes within 25 \%
for various sets of ${\hat c}_i$. The exception is provided by the
case in which the contribution of the $\Delta$ is eliminated from
${\hat c}_i$, and the $\Delta$ excitation current is taken into
account explicitly. Then the value of ${\hat d}^R$ is suppressed by
the factor \mbox{$\approx$ 4}.

Comparing our results with those of Ref.\,\cite{APKM} we see that
our calculations provide the value of
$\Lambda^0_{1/2}=254\,\rm{s}^{-1}$ which is by $\approx$ 4 \% larger
than the analogous value $\Gamma^{L=0}_{\mu d}=245\,\rm{s}^{-1}$ of
\cite{APKM}. Equally, our total capture rate
$\Lambda_{1/2}\,=416\,\rm{s}^{-1}$ differs from $\Gamma_{\mu
d}=386\,\rm{s}^{-1}$ \cite{APKM} by $\approx$ 7 \%.

In conclusion we stress that the planned precise experimental
investigation \cite{MSE} of the reaction (\ref{MUD}) is of
fundamental importance. It will stimulate efforts to understand
better the details and limits of application of both the TA and EFT
approaches and will certainly shed more light on the value of the
important LEC $\hat d^R$ (c$_D$).

\section*{Acknowledgments}
This work was partially supported by the grant GA \v{C}R 202/06/0746
and by Ministero dell' Istruzione, dell' Universit\`a e della
Ricerca of Italy (PRIN 2006). We thank Dr.\, Ji\v{r}\'{\i} Adam for
discussions and critical reading of the manuscript. The
correspondence with Dr. Doron Gazit is acknowledged.

\newpage

\appendix

\section{The weak currents of the TAA}
\label{appA}

The hadron currents consists of the one- and two-nucleon parts. The
one-nucleon currents are of the form, \bea {\vec
j}^{\,\,a}\,&=&\,\frac{1}{2M}\,[F^V_1(q^2)\,\P\,+\,iG^V_M(q^2)
(\vec \sigma\times\vq)]\,\frac{\tau^a}{2}\,,  \label{ONCVS}\\
\rho^{\,a}\,&=&\,F^V_1(q^2)\,\frac{\tau^a}{2}\,,  \label{ONCVT} \\ {\vec
j}^{\,\,a}_5\,&=&\,\left\{g_A F_A(q^2) \bigg[\vec \sigma\,-\,\frac{1}{8M^2}
[\P^{\,\,2}\vec\sigma\,-\,(\vec \sigma\cdot\P)\P\,+\,(\vec
\sigma\cdot\vq)\vq -i(\P\times\vq)]\bigg]  \right. \nonumber \\
&&\left. \,-\,\frac{g_{P}}{2M m_\mu}\,(\vec
\sigma\cdot\vq)\,\vq\,\right\}\,\frac{\tau^a}{2}\,, \label{ONCAS}\\
\rho^{\,a}_5\,&=&\,\bigg[\frac{g_A F_A(q^2)}{2M}(\vec
\sigma\cdot\P)\,-\,\frac{g_{P}}{2M m_\mu}(\vec
\sigma\cdot\vq)\,q_0\,\bigg]\frac{\tau^a}{2}\,. \label{ONCAT}  \eea
Here $\vec P=\vec p^{\,\,\prime}+\vec p$,
$q_\mu=p^{\,\prime}_{\,\mu}-p_{\,\mu}$, where $p^{\,\prime}_{\,\mu}$
($p_{\,\mu}$) is the four-momentum of the nucleon in the final
(initial) state and the induced pseudoscalar form factor is
\be g_P(\vec q^{\,\,2})= 2 M g_A m_\mu \Delta^\pi_F(\vec q^{\,\,2})\,.
\label{GP} \ee
For the other weak form factors we employ the dipole parametrization,
\bea
F^V_1(q^2)\,&=&\,1/(1\,+\,q^2/M_V^2)\,,\quad M_V^2\,=\,0.711\,\,\rm{GeV}^2\,, \label{WVFF} \\
F_A(q^2)\,&=&\,1/(1\,+\,q^2/M_A^2)\,,\quad M_A^2\,=\,1.04\,\,\rm{GeV}^2\,, \label{WAFF}
\eea
We use for the constant $g_A$ the value \cite{PDG}
\be g_A\,=\,-1.2694\,\pm\,0.0028\,. \label{GA} \ee

\subsection{The weak exchange currents}
\label{taawec}

The two-nucleon part also consists of the weak vector and weak axial
vector parts. We present first the weak vector MECs. They are
\begin{enumerate}
\item The $\pi$-pair term, \be \vec j^{\,\,a}(p.t.)\,=\,-\bigg(\frac{f_{\pi
NN}}{m_\pi}\bigg)^2\,F^V_1(q^2)\,\Delta^\pi_F(\qb^{\,\,2})\,F^2_{\pi
NN}(\qb^{\,\,2})\,\sa\,\s2q2\,\tta\,+\,\ot\,. \label{VPPT} \ee
\item
The pion-in-flight term, \bea \vec
j^{\,\,a}(\pi\pi)\,&=&\,\bigg(\frac{f_{\pi NN}}{m_\pi}\bigg)^2\,
\,F^V_1(q^2)\,\qa\,(\sa\cdot\qa)\,\s2q2\,\frac{1}{\qa^{\,\,2}-\qb^{\,\,2}}\,
\bigg[\Delta^\pi_F(\qb^{\,\,2})\,F^2_{\pi NN}(\qb^{\,\,2}) \bigg.
\nonumber \\ && \bigg. \,-\,\Delta^\pi_F(\qa^{\,\,2})\,F^2_{\pi
NN}(\qa^{\,\,2})\bigg]\,\tta\,+\,\ot\,.  \label{VPFT}  \eea
\item The $\Delta$ excitation current of the $\pi$ range,
\bea \vec j^{\,\,a}_\pi(\Delta)\,&=&\,-i\,\frac{q\,C^V_\pi}{9(M_\Delta-M)}\,F^V_1(q^2)\,
\hat q\,\times\,\left\{\,4\bigg[1\,+\,f(Y,Z)\bigg]\,\qb\,\tba
\right. \nonumber \\ && \left. \,+\,\bigg[1\,-\,2f(Y,Z)\bigg]\,
i(\sa\times\qb)\,\tta\,\right\} \nonumber \\
&&\,\times \Delta^\pi_F(\qb^{\,\,2})\,F^2_{\pi
NN}(\qb^{\,\,2})\,\s2q2\,+\,\ot\,, \label{VDCPIR} \eea where $\hat
q\,=\,\vq/|\vq|$ and
\bea f(Y,Z)\,&=&\,(1-M/M_\Delta)\,\left[C(Y)+C(Z)+2C(Y)C(Z)(2+M/M_\Delta)\right]\,, \nonumber \\
C(a)\,&=&\,-(\fot+a)\,,\quad C^V_\pi\,=\,2G_1\frac{f_{\pi N\Delta}\,f_{\pi NN}}{M\,m^2_\pi}\,.
\label{VDCPIRa} \eea
\item The $\Delta$ excitation current of the $\rho$ range,
\bea \vec j^{\,\,a}_\rho(\Delta)\,&=&\,i\,\frac{q\,C^V_\rho}{9(M_\Delta-M)}\,F^V_1(q^2)\,
\hat q\,\times\,\left\{\,4\bigg[1\,-\,2f(Y,Y)\bigg]\,(\qb\times(\sb\times\qb))\,\tba
\right. \nonumber \\ && \left. \,+\,\bigg[1\,+\,4f(Y,Y)\bigg]\,
i(\sa\times(\qb\times(\sb\times\qb)))\,\tta\,\right\} \nonumber \\
&&\,\times \Delta^\rho_F(\qb^{\,\,2})\,F^2_{\rho
NN}(\qb^{\,\,2})\,+\,\ot\,, \label{VDCRHOR} \eea where \be
C^V_\rho\,=\,\frac{1+\kappa^V_\rho}{2M}\,\bigg(\frac{g_\rho\, G_1}{
M}\bigg)^{\,2}\,. \label{VDCRHORa} \ee
\end{enumerate}
The current $\vec j^{\,\,a}(\pi\pi)$ of Eq.\,(\ref{VPFT}) is written
in such a form \cite{GR,ATA} that the potential current \be \vec
j^{\,\,a}(p.c.)\,=\,\vec j^{\,\,a}(p.t.)\,+\, \vec j^{\,\,a}(\pi\pi)
\label{VPC} \ee satisfies the CVC equation (\ref{NCVC}) for any form
factor $F_{\pi NN}$, if the one-pion exchange potential also
contains it. Here we use for the Gaussian form factor (\ref{EFF})
the following approximation in the current (\ref{VPFT}), \bea
\frac{1}{\qa^{\,\,2}-\qb^{\,\,2}}
\bigg[\Delta^\pi_F(\qb^{\,\,2})F^2_{\pi
NN}(\qb^{\,\,2})-\Delta^\pi_F(\qa^{\,\,2})F^2_{\pi
NN}(\qa^{\,\,2})\bigg]&=&\frac{1}{\qa^{\,\,2}-\qb^{\,\,2}}\Delta^\pi_F(\qa^{\,\,2})\Delta^\pi_F(\qb^{\,\,2})
F^2_{\pi NN}(\qb^{\,\,2}) \nonumber \\
&& \hspace{-8.6cm}\left\{(\qa^{\,\,2}-\qb^{\,\,2})+(\qb^{\,\,2}+m^2_\pi)
\left[1-e^{-(\qa^{\,\,2}-\qb^{\,\,2})/\Lambda^2_\pi}\right]\right\}=\Delta^\pi_F(\qa^{\,\,2})F^2_{\pi
NN}(\qb^{\,\,2})\left\{\Delta^\pi_F(\qb^{\,\,2})-\frac{1}{\Lambda^2_\pi}  \right. \nonumber \\
&&\left.\hspace{-8.8cm}\sum^\infty_{n=1} \frac{(-1)^n}{n!}
\left[\frac{\qa^{\,\,2}-\qb^{\,\,2}}{\Lambda^2_\pi}\right]^{n-1}\right\}\approx
\Delta^\pi_F(\qa^{\,\,2})F^2_{\pi NN}(\qb^{\,\,2})\bigg[\Delta^\pi_F(\qb^{\,\,2})
+\frac{1}{\Lambda^{2}_\pi}\bigg(1-\frac{\qa^{\,\,2}-\qb^{\,\,2}}{2\Lambda^2_\pi}\bigg)\bigg]\,. \label{APP1}
\eea

The weak axial MECs are
\begin{enumerate}
\item The $\pi$ potential current \cite{MRT2}, \bea \vec
j^{\,\,a}_{5\pi}(p.c.)\,&=&\,\bigg(\frac{f_{\pi
NN}}{ m_\pi}\bigg)^2\,\frac{g_A}{2M}\,
F_A(q^2)\, \large[\,(\vq + i \sa \times \Pa) \tba \,+\,(\Pa + i
\sa \times \vq) \tta
\large\,]  \nonumber \\
&&\,\times \Delta^\pi_F(\qb^{\,\,2})\,F^2_{\pi NN}(\qb^{\,\,2})\,
(\sb \cdot \qb)\,+\,\ot\,.  \label{APIPT} \eea
\item The
$\rho$-$\pi$ current, \bea \vec
j^{\,\,a}_{5}(\rho\pi)\,&=&\,-\bigg(\frac{f_{\pi
NN}}{ m_\pi}\bigg)^2\,\frac{1}{4Mg_A}\,
\large[\,1+\,m^2_\rho\,\Delta^\rho_F(\qa^{\,\,2})\,\large]\,\large[\,\Pa+
(1+\kappa^V_\rho)\,i\,(\sa \times \qa)\,\large] \nonumber \\
&&\,\times F_{\rho
NN}(\qa^{\,\,2})\,\Delta^\pi_F(\qb^{\,\,2})\,F_{\pi
NN}(\qb^{\,\,2})\, \,(\sb \cdot \qb)\tta\,+\,\ot \nonumber \\
\,&\approx&\,\bigg(\frac{f_{\pi NN}}{ m_\pi}\bigg)^2\,\frac{1}{4Mg_A}
(1+\kappa^V_\rho)\large[\,1+\,m^2_\rho\,\Delta^\rho_F(\qa^{\,\,2})\,\large]\,
i\,(\sa \times \qb) \nonumber \\
&&\,\times F_{\rho
NN}(\qa^{\,\,2})\,\Delta^\pi_F(\qb^{\,\,2})\,F_{\pi
NN}(\qb^{\,\,2})\, \,(\sb \cdot \qb)\tta\,+\,\ot\,. \label{RHPIT}
\eea Only the second part of Eq.\,(\ref{RHPIT}) contributes to the
rate $\Lambda_{1/2}$ sensibly.
\item The $\Delta$ excitation current of the
pion range, \bea \vec
j^{\,\,a}_{5\pi}(\Delta)\,&=&\,\frac{g_A}{9(M_\Delta-M)}\,
\bigg(\frac{f_{\pi N \Delta}}{m_\pi}\bigg)^2\,F_A(q^2)  \nonumber \\
&&\,\times [1\,-\,\vq\Delta^\pi_F(q^2)\,\vq\cdot\,]
\left\{\,4\left[1\,-\,\fot f(Z,Z)\right]\,\qb\tba\, \right. \nonumber \\
&&\left.\,+\,\left[1\,+\,f(Z,Z)\right]\,i\,(\sa\times \qb)\tta\,\right\} \nonumber \\
&&\,\times \Delta^\pi_F(\qb^{\,\,2})\,F^2_{\pi NN}(\qb^{\,\,2})\,
(\sb \cdot \qb)\,+\,\ot\,.  \label{ADCPIR} \eea
\item The $\Delta$
excitation current of the $\rho$ meson range,
\bea \vec j^{\,\,a}_{5\rho}(\Delta)\,&=&\,\frac{g_A\,C^A_\rho}{9(M_\Delta-M)}\,F_A(q^2) \nonumber \\
&&\,\times [1\,-\,\vq\Delta^\pi_F(q^2)\,\vq\cdot\,]
\left\{\,4\left[1\,+\,f(Y,Z)\right]
\,\qb\times(\sb\times \qb)\tba \right. \nonumber \\
&&\left.\,+\,\left[1\,-\,2f(Y,Z)\right]\,i\sa\times(\qb\times(\sb\times
\qb))\tta\,\right\} \nonumber \\
&&\,\times \Delta^\rho_F(\qb^{\,\,2})\,F^2_{\rho
NN}(\qb^{\,\,2})\,+\,\ot\,,  \label{ADCRHOR} \eea where \be
C^A_\rho\,=\,G_1\bigg(\frac{ g_\rho}{M}\bigg)^2\,
\frac{1+\kappa^V_\rho}{4}\,\frac{f_{\pi N\Delta}}{f_{\pi NN}}\,.
\label{ADCRHORa} \ee
\item The potential current of the $\rho$ range \cite{MRT1}
\bea \vec
j^{\,\,a}_{5\rho}(p.c.)\,&=&\,\left(\frac{g_\rho}{2}\right)^2
\frac{(1+\kappa^V_\rho)^2}{(2M)^3}\,g_A F_A(q^2)\, \left\{\tau^a_2
\left[\vec q\times(\vec \sigma_2\times \vec q_2) + i\vec
\sigma_1\times (\vec P_1 \times
(\vec \sigma_2\times \vec q_2))\right] \right. \nonumber  \\
&& \left. \,+i(\vec \tau_1\times\vec \tau_2)^a \left[\vec P_1\times
(\vec \sigma_2\times \vec q_2) +i\vec \sigma_1\times(\vec q\times
(\vec \sigma_2\times \vec q_2))\right]\right\} \nonumber
\\ &&\,\times \Delta^\rho_F(\vec q^{\,\,2}_2)
\,F^2_{\rho NN}(\qb^{\,\,2})\, \nonumber \\
&&\,-\left(\frac{g_\rho}{2}\right)^2\frac{(1+\kappa^V_\rho)}{(2M)^2}\,
\frac{g_P({\vec q}^{\,\,2})}{m_l}\,\vec q\,\left[\tau^a_2(\vec \sigma_1
\cdot \vec q_2)\,+i(\vec \tau_1\times\vec \tau_2)^a\,i(\vec
\sigma_1\cdot \vec \sigma_2 \times \vec q_2)\right]
\nonumber \\
&&\,\times \Delta^\rho_F(\vec q^{\,\,2}_2)\,F^2_{\rho
NN}(\qb^{\,\,2})\, +\,\ot\,, \label{APCRHOR} \eea
\item The time component of the weak axial MEC
\bea \rho^a_5(\rho\pi)\,&=&\,-\bigg(\frac{f_{\pi
NN}}{ m_\pi}\bigg)^2\,\frac{1}{2g_A}\,
\large[\,1+\,m^2_\rho\,\Delta^\rho_F(\qa^{\,\,2})\,\large]
\,F_{\rho NN}(\qa^{\,\,2})\,\Delta^\pi_F(\qb^{\,\,2})\,F_{\pi
NN}(\qb^{\,\,2}) \nonumber\\
&&\,\times (\sb \cdot \qb)\tta\,+\,\ot\,. \label{TCAMEC} \eea
\end{enumerate}
In Eqs.\,(\ref{VPPT})-(\ref{TCAMEC}),  $\vec P_i=\vec p^{\,\,\prime}_i+\vec p_i$,
$\vq = \qa + \qb$, $\vec q_i =\vec p^{\,\,\prime}_i-\vec p_i$. In reaction (\ref{MUD}),
the isospin components of the currents and charge densities ${\cal O}^{\,\,-}\,=
\,{\cal O}^{\,\,1}\,-i\,{\cal O}{\,\,^2}$ are effective.

\subsection{The Fourier transform of the weak MECs}
\label{fttaawec}

Here we provide the Fourier transform of the weak MECs in the same
order as they are listed in the previous section. We start with
presenting the form factors arising due to the form factors of the
type (\ref{EFF}) after the Fourier transformation of the weak MECs
containing one boson propagator. They are \bea
W_{0\,B}\,&\equiv&\,\phi^0_c(x_B)\,=\,e^{(m_B/\Lambda_B)^2}\left[e^{-x_B}\,erfc\bigg(-\frac{\Lambda_B
x_B} {2
m_B}+\frac{m_B}{\Lambda_B}\bigg)\,-\,e^{x_B}\,erfc\bigg(\frac{\Lambda_B
x_B}
{2 m_B}+\frac{m_B}{\Lambda_B}\bigg)\right] \nonumber\\
&&\,/(2 x_B)\,,  \label{W0B} \\
W_{1\,B}\,&\equiv&\,-\frac{d \phi^0_c(x_B)}{d x_B}\,=\,\left\{\phi^0_c+e^{(m_B/\Lambda_B)^2}\left[e^{-x_B}\,erfc\bigg(-\frac{\Lambda_B x_B}
{2 m_B}+\frac{m_B}{\Lambda_B}\bigg)  \right.\right. \nonumber \\
&&\left.\left.\,+\,e^{x_B}\,erfc\bigg(\frac{\Lambda_B x_B}
{2 m_B}+\frac{m_B}{\Lambda_B}\bigg)\right] /2-\frac{\Lambda_B}{\sqrt{\pi} m_B}\,e^{-(\Lambda_B x_B/2 m_B)^2}\right\}/x_B\,,
\label{W1B} \\
W_{2\,B}\,&\equiv&\,\frac{d^2 \phi^0_c}{d x^2_B}-\frac{d \phi^0_c}{x_B d x_B}\,=\,
\bigg(1+\frac{3}{x^2_B}\bigg)\phi^0_c+\frac{3}{2x^2_B}e^{(m_B/\Lambda_B)^2}\left[e^{-x_B}\,erfc\bigg(-\frac{\Lambda_B x_B}
{2 m_B}+\frac{m_B}{\Lambda_B}\bigg)  \right. \nonumber \\
&&\left.\,+\,e^{x_B}\,erfc\bigg(\frac{\Lambda_B x_B}
{2 m_B}+\frac{m_B}{\Lambda_B}\bigg)\right]-\frac{\Lambda_B}{2\sqrt{\pi} m_B}\,\bigg[\bigg(\frac{\Lambda_B}
{m_B}\bigg)^2+\frac{6}{x_B^2}\bigg]e^{-(\Lambda_B x_B/2 m_B)^2}\,,\label{W2B} \\
W_B\,&=&\,\frac{1}{2\sqrt{\pi}}\bigg(\frac{\Lambda_B}{m_B}\bigg)^3\, e^{-(\Lambda_B x_B/2 m_B)^2}\,, \label{WB}\\
W_{2\,B}\,&=&\,W_{0\,B}\,+\,\frac{3}{x_B}\,W_{1\,B}\,-\,W_B\,. \label{WREL}
\eea
In  Eqs.\,(\ref{W0B})-(\ref{W2B}),  the function $erfc(x)$ is
the complementary error function \cite{AS}.

First follow the weak vector MECs:
\begin{enumerate}
\item The $\pi$-pair term, \be {\vec {\tilde j}}^{\,\,a}(p.t.)\,=\,-i\frac{f^2_{\pi
NN}}{4\pi}\,F^V_1(q^2)\,\sa(\sb \cdot \hat r)\,\tta\, e^{i(\vq\cdot \vec r_1)}\,W_{1\pi}\,+\,\ot\,.
\label{FTVPPT} \ee
\item The pion-in-flight term,
\bea {\vec{\tilde j}}^{\,\,a}(\pi\pi)\,\equiv\,\sum_{i=1}^4\,{\vec{\tilde j}}^{\,\,a}_i(\pi\pi)\,&=&\,
\frac{1}{2\pi^{3/2}\,q}\,\bigg(\frac{f_{\pi NN}}{m_\pi}\bigg)^2\,
\,F^V_1(q^2)\,
\tta\,e^{i(\vq\cdot \vec r_1)} \nonumber\\
&&\,\times \sum^4_{i=1} {\vec{\cal O}}_i\,f^0_{LL}(r)\,+\,\ot\,,
\label{FTVPFT} \eea where \bea {\vec{\cal
O}}_1\,&=&\,-i\vq(\sa\cdot\vq)(\sb\cdot\nabla_{\vec r})\,,\quad\quad
{\vec{\cal O}}_2\,=\,\vq(\sa\cdot\nabla_{\vec r})(\sb\cdot\nabla_{\vec r})\,, \nonumber \\
{\vec{\cal O}}_3\,&=&\,(\sa\cdot\vq)(\sb\cdot\nabla_{\vec r})\nabla_{\vec r}\,,\quad\,\,\,\,
{\vec{\cal O}}_4\,=\,i(\sa\cdot\nabla_{\vec r})(\sb\cdot\nabla_{\vec r})\nabla_{\vec r}\,,\label{OI}
\eea
and
\bea
f^0_{LL}(r)\,&=&\,\sum_L\,i^L\,Y_{L0}(\hat r){\hat L}F^0_{LL}\,,  \label{FR0LL} \\
F^n_{LK}\,&=&\,\int^{+\infty}_0\,dp\,p^{(1+n)}\,
e^{-(p/\Lambda_\pi)^2}\,j_K(pr)\,\left\{Q_L(\beta)\left[
\frac{1}{p^2+m^2_\pi} \right.\right. \nonumber \\
&&\,\left.\left. +\frac{1}{\Lambda_\pi^2}\left(1+\frac{p^2+m^2_\pi}{2\Lambda^2_\pi}\right)\right]
-\delta_{L0}\frac{p q}{\Lambda^4_\pi}\right\}\,.   \label{FNLK}
\eea
Here \mbox{$\beta=(p^2+q^2+m^2_\pi)/2pq$} and $Q_L(\beta)$ is the Legendre polynomial of the second sort \cite{AS}.

Numerically, only the current ${\vec{\tilde j}}^{\,\,a}_3(\pi\pi)$ contributes non-negligibly to the $\Lambda_{1/2}$
for the transition $d\rightarrow {1}S_0$,
\bea
{\vec{\tilde j}}^{\,\,a}_3(\pi\pi)\,&=&\,-\frac{1}{\pi \sqrt{3}}\,\bigg(\frac{f_{\pi NN}}{m_\pi}\bigg)^2\,
\,F^V_1(q^2)\,\tta\,e^{i(\vq\cdot \vec r_1)}\sum_{L}\,i^{L}\,\sum_{N=L\pm 1}\,c_N \nonumber \\
\,&&\,\times \sum_{K=N\pm 1}d_K F^2_{L K}\,\sum_{j g
h}\,(-1)^g\hat{j}\hat g\hat h\seij{L}{1}{N}{K}{1}{j}
\seij{K}{1}{j}{h}{g}{1} \nonumber \\
\,&&\,\times [\hat e \otimes [ [\sa\otimes\sb]^h \otimes [Y_1(\hat
q)\otimes Y_K(\hat r)]^g]^j]^{L0}\,+\ot\,, \label{FTJPP3} \eea where
$\hat e=\hat e_{\pm},\hat e_0$ are the orthogonal unit vectors, and
\be
c_{L+1}=\sqrt{L+1}\,,\,c_{L-1}=\sqrt{L}\,,\,d_{N+1}=\sqrt{N+1}\,,\,d_{N-1}=\sqrt{N}\,.
\label{CDS} \ee
\item The $\Delta$ excitation current of the $\pi$ range,
\bea {\vec {\tilde j}}^{\,\,a}_\pi(\Delta)&=&-i\frac{q\,C^V_\pi m^3_\pi}{36\pi (M_\Delta-M)}F^V_1(q^2)
e^{i(\vq\cdot \vec r_1)}\,\sum_{i=1}^4\,F^V_{i\,\pi}{\hat q}\times{\vec{\cal O}}^{a}_i(\Delta)\,+\,\ot\,, \label{FTVDCPIR} \\
{\vec{\cal O}}^{a}_1(\Delta)\,&=&\,\sb\,\tba\,, \label{OV1P}\\
{\vec{\cal O}}^{a}_2(\Delta)\,&=&\,-{\hat r}\,(\sb\cdot\hat r)\tba
\,, \label{OV2P}\\
{\vec{\cal O}}^{a}_3(\Delta)\,&=&\,i(\sa\times\sb)\,\tta\,, \label{OV3P} \\
{\vec{\cal O}}^{a}_4(\Delta)\,&=&\,i(\hat r\times\sa)(\sb\cdot\hat r)\,\tta
\,, \label{OV4P} \\
F^V_{1\, \pi}\,&=&\,\frac{4}{x_\pi}\,W_{1\,\pi}\,[1+f(Y,Z)]\,,\quad
F^V_{2\, \pi}\,=\,4\,W_{2\,\pi}\,[1+f(Y,Z)]\,, \nonumber \\
F^V_{3\,\pi}\,&=&\,\frac{1}{x_\pi}\,W_{1\,\pi}\,[1-2f(Y,Z)]\,,\quad
F^V_{4 \,\pi}\,=\,W_{2\,\pi}\,[1-2f(Y,Z)]\,. \label{FVISP}
\eea
\item The $\Delta$ excitation current of the $\rho$ range,
\bea
{\vec {\tilde j}}^{\,\,a}_\rho(\Delta)\,&=&\,-i\,\frac{q\,C^V_\rho
m^3_\rho}{36\pi (M_\Delta-M)}\,F^V_1(q^2)e^{i(\vq\cdot \vec r_1)}\,
\sum_{i=1}^4\,F^V_{i\,\rho}\,{\hat q}\times{\vec{\cal O}}^{a}_i(\Delta)\,+\,\ot\,, \label{FTVDCRHOR} \\
F^V_{1\,\rho}\,&=&\,4(W_{2\,\rho}-\frac{2}{x_\rho}\,W_{1\,\rho})\,[1-2f(Y,Y)]\,,\,\,
F^V_{2\,\rho}\,=\,4\,W_{2\,\rho}\,[1-2f(Y,Y)]\,, \nonumber \\
F^V_{3\,\rho}\,&=&\,(W_{2\,\rho}-\frac{2}{x_\rho}\,W_{1\,\rho})\,[1+4f(Y,Y)]\,,\,\,
F^V_{4\,\rho}\,=\,W_{2\,\rho}\,[1+4f(Y,Y)]\,. \label{FVISR}
\eea
\end{enumerate}
Next we present the Fourier transform of the weak axial MECs:
\begin{enumerate}
\item The $\pi$ potential current,
\bea {\vec{\tilde j}}^{\,\,a}_{5\pi}(p.c.)\,&=&\,\frac{f^2_{\pi NN}}{4\pi}\frac{m_\pi}{2M}g_A\,
F_A(q^2)\,e^{i(\vq\cdot \vec r_1)}
\sum_{i=1}^{10}\,F_{i\pi}(p.c.)\,{\vec{\cal O}}^a_{i\pi}(p.c.)\,+\ot\,,   \label{FTAPIPT} \\
{\vec{\cal O}}^a_{1\pi}(p.c.)\,&=&\,(\vq\times\sa)\,(\sb\cdot\hat r)\tba\,,\quad
{\vec{\cal O}}^a_{2\pi}(p.c.)\,=\,2i\,(\sb\cdot\hat r)(\sa\times\nabla_1)\tba  \,,\label{O12PC} \\
{\vec{\cal O}}^a_{4\pi}(p.c.)\,&=&\, i\vq(\sb\cdot\hat r)\tta  \,,\,
{\vec{\cal O}}^a_{5\pi}(p.c.)\,=\,2(\sb\cdot\hat r){\nabla_1}\tta  \,,\,\label{O45PC} \\
{\vec{\cal O}}^a_{3\pi}(p.c.)\,&=&\,i\vq(\sb\cdot\hat r)\tba\,,\quad
{\vec{\cal O}}^a_{6\pi}(p.c.)=(\sa\times\vq)(\sb\cdot\hat r)\tta   \,,\label{O36PC} \\
{\vec{\cal O}}^a_{7\pi}(p.c.)\,&=&\,-i(\sa\times\sb)\tba  \,,\quad
{\vec{\cal O}}^a_{8\pi}(p.c.)\,=\,-\sb\tta   \,,\label{O78PC} \\
{\vec{\cal O}}^a_{9\pi}(p.c.)\,&=&\,i(\sa\times\hat r)(\sb\cdot\hat r)\tba   \,,\quad
{\vec{\cal O}}^a_{10\pi}(p.c.)\,=\,{\hat r}(\sb\cdot\hat r)\tta  \,,\label{O910PC} \\
F_{1\pi}(p.c.)\,&=&\,F_{2\pi}(p.c.)=F_{3\pi}(p.c.)=F_{4\pi}(p.c.)=F_{5\pi}(p.c.)=F_{6\pi}(p.c.) \nonumber \\
\,&=&\,W_{1\pi}/m_\pi\,, \label{F1_6PC} \\
F_{7\pi}(p.c.)\,&=&\,F_{8\pi}(p.c.)=W_{1\pi}/x_\pi\,, \label{F78PC} \\
F_{9\pi}(p.c.)\,&=&\,F_{10\pi}(p.c.)=W_{2\pi}\,. \label{F910PC}
\eea

\item The $\rho$-$\pi$ current,
\bea {\vec{\tilde j}}^{\,\,a}_{5}(\rho\pi)\,&=&\,\bigg(\frac{f_{\pi
NN}}{ m_\pi}\bigg)^2\,\frac{1}{4Mg_A}\,\frac{1}{\pi q}\sqrt{\frac{3}{2\pi}}\,
e^{-\frac{q^2}{2\Lambda^2_\rho}(1-a)+ia(\vq\cdot \vec r)+i(\vq\cdot \vec r_1)}\,i(1+\kappa^V_\rho)\nonumber \\
\,&&\,\times \tta[\,\Delta_1 {\vec{\tilde j}}^{\,\,a}_{5}(\rho\pi)
+\,\Delta_2 {\vec{\tilde j}}^{\,\,a}_{5}(\rho\pi)\,]
\,+\,\ot\,.\label{FTRHPIT} \eea Here \bea \Delta_1 {\vec{\tilde
j}}^{\,\,a}_{5}(\rho\pi)\,&=&\,m^2_\rho\,
\sum_L\,i^{L+1}\bigg<\,\sqrt{L+1}\sum_{N=L,L+2}\,a_N H^3_{L N}(r)\sum_{j k}\,(-1)^k{\hat j}{\hat k} \bigg.\nonumber \\
\,&&\,\bigg. \times \novej{1}{N}{L+1}{1}{k}{L}{1}{j}{1}\,[{\cal
C}(j,k,N)]^{L0}
\,+\,\sqrt{L}\sum_{N=L-2,L}b_N  \bigg.\nonumber \\
\,&&\,\bigg. \times H^3_{L N}(r)\,\sum_{j k}\,(-1)^k{\hat j}{\hat
k}\, \novej{1}{N}{L-1}{1}{k}{L}{1}{j}{1}\,[{\cal
C}(j,k,N)]^{L0}\,\bigg>\,, \label{D1TJRHOPI} \eea where \bea
a\,&=&\,\frac{\Lambda^2_{\pi\rho}}{\Lambda^2_\rho}\,,\quad
\frac{1}{\Lambda^2_{\pi\rho}}\,\equiv\,
\frac{1}{\Lambda^2_\pi}\,+\,\frac{1}{\Lambda^2_\rho}\,, \label{A} \\
a_L\,&=&\,\sqrt{L+1}\,,\quad a_{L+2}\,=\,\sqrt{L+2}\,,\quad b_{L-2}\,=\,\sqrt{L-1}\,,\quad
b_{L}\,=\,\sqrt{L}\,,\label{ALBL} \\
H^n_{L N}(r)\,&=&\,\int^{+\infty}_0\,dp\, p^n\,e^{-\frac{p^2}{2a\Lambda^2_\rho}}\,j_N(pr)F_L(p,q)\,, \label{ILN} \\
F_L(p,q)\,&=&\,\frac{Q_L(\alpha)\,-\,Q_L(\beta)}{p^2+a(1-a)\,q^2+(1-a)\,m^2_\pi+a\,m^2_\rho}\,,  \label{FLPQ} \\
\alpha\,&=&\,(p^2+a^2\,q^2+m^2_\pi)/(2apq)\,,\quad  \beta\,=\,[p^2+(1-a)^2\,q^2+m^2_\rho] \nonumber \\
\,&&\,/[2(a-1)pq]\,. \label{ALPBE}
\eea
Further the symbols $[{\cal C}(j,k,N)]^{L0}$ are defined as
\be [{\cal C}(j,k,N)]^{L0}\,=\,[\hat e \otimes [Y_N(\hat r)\otimes[\sa\otimes\sb]^j]^k]^{L0}\,,
\label{CC} \ee
and the piece
$\Delta_2 {\vec{\tilde j}}^{\,\,a}_{5}(\rho\pi)$ can be obtained from the term
$\Delta_1 {\vec{\tilde j}}^{\,\,a}_{5}(\rho\pi)$, Eq.\,(\ref{D1TJRHOPI}), by the change
\be m^2_\rho\,\rightarrow\,1/a\,, \quad F_L(p,q)\,\rightarrow\,Q_L(\alpha)\,. \label{CHANGE}
\ee
At last, the symbols $\novej{a_1}{b_1}{c_1}{a_2}{b_2}{c_2}{a_3}{b_3}{c_3}$ in Eq.\,(\ref{D1TJRHOPI})
are Wigner's 9j symbols \cite{VMK}.
\item The $\Delta$ excitation current of the $\pi$ range,
\bea {\vec {\tilde j}}^{\,\,a}_{5\pi}(\Delta)&=&\frac{g_A F_A
m^3_\pi}{36 \pi(M_\Delta-M)}
\bigg(\frac{f_{\pi N \Delta}}{m_\pi}\bigg)^2 e^{i(\vq\cdot \vec r_1)} \nonumber \\
&& \times [1-\vq\Delta^\pi_F(q^2)\vq\cdot]
\,\sum_{i=1}^4\,F^A_{i\,\pi}\,{\vec{\cal O}}^{a}_i(\Delta)\,+\,\ot\,,\label{FTADCPIR}\\
F^A_{1\, \pi}\,&=&\,\frac{4}{x_\pi}\,W_{1\,\pi}\,[1-\fot
f(Z,Z)]\,,\quad
F^A_{2\, \pi}\,=\,4\,W_{2\,\pi}\,[1-\fot f(Z,Z)]\,, \nonumber \\
F^A_{3\,\pi}\,&=&\,\frac{1}{x_\pi}\,W_{1\,\pi}\,[1+f(Z,Z)]\,,\quad
F^A_{4 \,\pi}\,=\,W_{2\,\pi}\,[1+f(Z,Z)]\,. \label{FAISP} \eea
\item The $\Delta$ excitation current of the $\rho$ range,
\bea {\vec {\tilde
j}}^{\,\,a}_{5\rho}(\Delta)\,&=&\,-\frac{g_A\,m^3_\rho
C^A_\rho}{36\pi (M_\Delta-M)}\,
F_A(q^2)e^{i(\vq\cdot \vec r_1)} \nonumber \\
&& \,\times [1-\vq\Delta^\pi_F(q^2)\vq\cdot]
\,\sum_{i=1}^4\,F^A_{i\,\rho}\,{\vec{\cal O}}^{a}_i(\Delta)\,+\,\ot\,, \label{FTADCRHOR} \\
F^A_{1\,\rho}\,&=&\,4(W_{2\,\rho}-\frac{2}{x_\rho}\,W_{1\,\rho})\,[1+f(Y,Z)]\,,\,\,
F^A_{2\,\rho}\,=\,4\,W_{2\,\rho}\,[1+f(Y,Z)]\,, \nonumber \\
F^A_{3\,\rho}\,&=&\,(W_{2\,\rho}-\frac{2}{x_\rho}\,W_{1\,\rho})\,[1-2f(Y,Z)]\,,\,\,
F^A_{4\,\rho}\,=\,W_{2\,\rho}\,[1-2f(Y,Z)]\,. \label{FAISR} \eea
\item The potential current of the $\rho$ range
\bea {\vec{\tilde j}}^{\,\,a}_{5\rho}(p.c.)\,&=&\,\frac{1}{4\pi}\left(\frac{g_\rho}{2}\right)^2
\left(\frac{m_\rho}{2M}\right)^3g_A F_A(q^2)(1+\kappa^V_\rho)^2e^{i(\vq\cdot \vec r_1)}
\sum_{i=1}^{6}F_{i\rho}(p.c.)  \nonumber \\
\,&&\,\times {\vec{\cal O}}^a_{i\rho}(p.c.)\,
+\,\ot\,,  \label{FTAPCRHOR} \\
{\vec{\cal O}}^a_{1\rho}(p.c.)&=&{q}(\sa\times\hat r)(\sb\cdot\hat q)\tba\,,\,
{\vec{\cal O}}^a_{2\rho}(p.c.)=-{2i}(\sa\times\hat r)(\sb\cdot\nabla_1)\tba\,,\label{O12RPC} \\
{\vec{\cal O}}^a_{3\rho}(p.c.)&=&-i(\sa\times\hat r)(\sb\cdot\hat r)\tba\,,\,
{\vec{\cal O}}^a_{4\rho}(p.c.)=-{iq}{\hat r}(\sb\cdot\hat q)\tta\,,\,\label{O34RPC} \\
{\vec{\cal O}}^a_{5\rho}(p.c.)&=&-{2}{\hat r}(\sb\cdot\nabla_1)\tta\,,\,
{\vec{\cal O}}^a_{6\rho}(p.c.)=-{\hat r}(\sb\cdot\hat r)\tta\,,\,\label{O56RPC} \\
F_{1\rho}(p.c.)\,&=&\,F_{2\rho}(p.c.)=F_{4\rho}(p.c.)=F_{5\rho}(p.c.)=W_{1\rho}\,, \label{F1245RPC} \\
F_{3\rho}(p.c.)\,&=&\,F_{6\rho}(p.c.)=m_\rho\,W_{2\rho}\,. \label{F36RPC}
\eea
\item The time component of the weak axial MEC
\bea {\tilde \rho}^a_5(\rho\pi)\,&=&\,-\bigg(\frac{f_{\pi
NN}}{ m_\pi}\bigg)^2\,\frac{1}{2g_A}\,\frac{a}{2\pi^{3/2}}
e^{-\frac{q^2}{2\Lambda^2_\rho}(1-a)+ia(\vq\cdot \vec r)+i(\vq\cdot \vec r_1)}\,(\sb\cdot \hat q) \nonumber \\
\,&&\,\times \tta[\Delta_1 {\tilde \rho}^a_5(\rho\pi)\,+\,\Delta_2
{\tilde \rho}^a_5(\rho\pi)]\,+\,\ot\,. \label{FTTCAMEC} \eea Here
\be \Delta_1 {\tilde \rho}^a_5(\rho\pi)\,=\,m^2_\rho\,
\sum_L\,i^{-L}{\hat L}\,Y_{L0}(\hat r)\,H^1_{LL}(r)\,.
\label{DEL1TC} \ee The function $H^n_{LN}(r)$ is defined in
Eq.\,(\ref{ILN}) and the piece $\Delta_2 {\tilde \rho}^a_5(\rho\pi)$
can be obtained from the term $\Delta_1 {\tilde
\rho}^{\,\,a}_{5}(\rho\pi)$, Eq.\,(\ref{DEL1TC}), by the change
(\ref{CHANGE}).
\end{enumerate}

\section{The EFT currents}
\label{appB}

For the one-body currents, we take the currents from
Eqs.\,(\ref{ONCVS})--(\ref{ONCAT}) of Appendix \ref{appA}, with the
form factors in the quadratic radius approximation \cite{KA,LI},
\bea
F^V_1(q^2)\,&\approx&\,1\,-\,\frac{1}{6}\,r^2_V\,q^2\,,\quad r^2_V\,=\,0.59\,\,\rm{fm}^2\,, \label{FV1QRA} \\
F_A(q^2)\,&\approx&\,1\,-\,\frac{1}{6}\,r^2_A\,q^2\,,\quad
r^2_A\,=\,(0.403\pm 0.030)\,\,\rm{fm}^2\,, \label{FAQRA} \eea
However, this approximation changes the results only a little bit in
comparison with the dipole form factors used in the TAA
calculations.

\subsection{The weak exchange currents}
\label{eftwec}

For the weak vector currents, we take the currents from
Eqs.\,(\ref{VPPT})--(\ref{VDCRHOR}) of Appendix \ref{appA}. So we
add to the $\pi$-pair and pion-in-flight terms considered in
\cite{APKM} the $\Delta$ excitation currents of the $\pi$ and $\rho$
ranges. Inspecting Table 2 of Ref.\,\cite{PMR} one can in addition
expect a non-negligible contribution also from the $\rho \omega \pi$
and two-pion exchange currents.

As to the weak axial MEC operator, we adopt here the main part of
this current used in \cite{APKM} and add to it the $\pi$ potential
current of Eq.\,(\ref{APIPT}) demanded by the PCAC constraint
(\ref{NPCAC}). In our notation, the currents (19)--(21) of
Ref.\,\cite{APKM} are: \bea
{\hat A}^0_{2B}\,&=&\,-\frac{g_A F_A}{4 f^2_\pi}\Delta^\pi_F(\qb^{\,\,2})\,\s2q2\,\tta\,+\,\ot\,, \label{A02B} \\
{\hat {\vec A}}_{2B}\,&=&\,\frac{g_A F_A}{2M f^2_\pi}\left\{\,\left[\frac{1}{4}\,\Pa\,\tta + 2{\hat c}_3\,\qb\tba
-({\hat c}_4+\frac{1}{4})i(\sa\times\qb)\,\tta \right.\right. \nonumber \\
\,&&\,\left.\left. +\frac{1+c_6}{4}\,i(\sa\times\vq)\,\tta\,\right]\,\Delta^\pi_F(\qb^{\,\,2})\,\s2q2
\right. \nonumber \\ \,&&\,\left. +\,{\hat d}_1(\sa\ta+\sb\tba)\,-\,{\hat d}_2\,i(\sa\times\sb)\,\tta\right\}
\,+\,\ot\,,    \label{A2BSC} \\
{\hat P}(\qa,\qb)\,&=&\,\frac{g_A F_A m^2_\pi}{2M f^2_\pi}\,4{\hat
c}_1\,\tba\,\Delta^\pi_F(\qb^{\,\,2})\,\s2q2 \,+\,\ot\,.  \label{HP}
\eea In Eq.\,(\ref{A02B}), we keep only the leading order of the
time component considered in Eq.\,(18) \cite{APKM}. because the
contribution of this part of ${\hat A}^0_{2B}$ is small and the
correction to it is suppressed by the factor $\approx$ 1/M$^2$. Let
us note that the time component (\ref{A02B}) is the soft pion
approximation to its hard pion form (\ref{TCAMEC}).

\subsection{The Fourier transform of the EFT weak axial MECs}
\label{ftwaeftmec} Multiplying the currents (\ref{A02B}),
(\ref{A2BSC}) and (\ref{HP}) by the form factor squared of the type
(\ref{EFF}) we obtain

\bea
{\tilde{\hat A}^0}_{2B}\,&=&\,-i\frac{1}{4\pi}\frac{g_A  F_A m^2_\pi}{4 f^2_\pi}(\sb\cdot\hat r)e^{i(\vq\cdot \vec r_1)}
\,W_{1\pi}\,\tta\,+\,\ot\,, \label{FTA02B} \\
{\tilde{\hat {\vec A}}}_{2B}\,&=&\,\frac{1}{4\pi}\frac{g_A F_A m^3_\pi}{2M f^2_\pi}e^{i(\vq\cdot \vec r_1)}\sum^{10}_{i=1}
\,F_{i}(2B)\,{\vec{\cal O}}^a_{i}(2B)\,+\ot\,,   \label{FTA2BSC} \\
{\vec{\cal O}}^a_{1}(2B)\,&=&\,i\vq\,(\sb\cdot\hat r)\tta\,,\quad
{\vec{\cal O}}^a_{2}(2B)\,=\,2\,(\sb\cdot\hat r)\,\nabla_1\tta  \,,\label{O122B} \\
{\vec{\cal O}}^a_{3}(2B)\,&=&\, -\sb\,\tta  \,,\quad
{\vec{\cal O}}^a_{4}(2B)\,=\,\hat r(\sb\cdot\hat r)\tta  \,,\,\label{O342B} \\
{\vec{\cal O}}^a_{5}(2B)\,&=&\,2\sb\,\tba\,,\quad
{\vec{\cal O}}^a_{6}(2B)=-2\hat r(\sb\cdot\hat r)\tba   \,,\label{O562B} \\
{\vec{\cal O}}^a_{7}(2B)\,&=&\,-i(\sa\times\sb)\tta  \,,\quad
{\vec{\cal O}}^a_{8}(2B)\,=\,i(\sa\times \hat r)(\sb\cdot\hat r)\tta   \,,\label{O782B} \\
{\vec{\cal O}}^a_{9}(2B)\,&=&\,-(\sa\times\vq)(\sb\cdot\hat r)\tta   \,,\quad
{\vec{\cal O}}^a_{10}(2B)\,=\,\sa\ta\,,\label{O902B}  \\
F_{1}(2B)\,&=&\,F_{2}(2B)\,=\,W_{1\pi}/4m_\pi\,,\quad F_{3}(2B)\,=\,W_{1\pi}/4 x_\pi\,,\quad
F_{4}(2B)\,=\,W_{2\pi}/4\,,  \nonumber \\
F_{5}(2B)\,&=&\,{\hat c}_3\,W_{1\pi}/x_\pi\,+\,{\hat d}_1 W_{\pi}/2\,,\quad F_{6}(2B)\,=\,{\hat c}_3\,W_{2\pi}\,, \nonumber \\
F_{7}(2B)\,&=&\,({\hat c}_4+\frac{1}{4})\,W_{1\pi}/x_\pi+{\hat d}_2
W_\pi\,,
 \quad F_{8}(2B)\,=\,({\hat c}_4+\frac{1}{4})\,W_{2\pi}\,,\nonumber \\
F_{9}(2B)\,&=&\,\frac{1+c_6}{4}\,W_{1\pi}/m_\pi\,,\quad F_{10}(2B)\,=\,{\hat d}_1\,W_\pi\,. \label{FI2B} \\
{\tilde{\hat P}}(\qa,\qb)\,&=&\,i\frac{1}{4\pi}\frac{2g_A F_A
m^2_\pi}{M f^2_\pi}\,{\hat c}_1\,(\sb\cdot\hat r)e^{i(\vq\cdot \vec
r_1)} \,W_{1\pi}\,\tba\,+\,\ot\,. \label{FTHP}
\eea

\section{The multipoles of the currents}
\label{appC}

We first present the contribution  to the multipoles from the IA
currents. In order to make the equations more transparent, we do not
write the argument $qr/2$ in the Bessel functions $j_i(qr/2)$,
unless the argument differs, which is the case of the $\rho$-$\pi$
current. The factor $\st$ arising from the isovector matrix elements
is not kept in the reduced matrix elements of the current, but is
included in the overall constants in front of the integrals in
Eqs.\,({\ref{LSTAT1}) and (\ref{DL}). We also take into account the
factor $1/\kappa_0$, entering the reduced matrix elements according
to the definition given in Eq.\,(4.20) of Ref.\,\cite{MRT1}, by
keeping $1/\kappa^2_0$ in the integration volume in the same
Eqs.\,({\ref{LSTAT1}) and (\ref{DL}).

\subsection{The multipoles of the IA currents}
\label{MIAC}

\subsubsection{J=0 multipoles}
\label{J0M}

\bea <^{3}P_1||{\hat L}_{\,50}||d>&=&i\st [-g_A F_A
(1-\frac{\vq^{\,\,2}}{8M^2})+\frac{g_P\,\vq^{\,\,2}}{2Mm_\mu}]\int^{+\infty}_0\,dr\,
u^1_{11,2}(\kappa,r)\, j_1\left(u_0(r) \right.\nonumber \\
&&\left. +u_2(r)/\st\right)\,+i\st \frac{g_A F_A}{M^2}\left<\frac{q}{6}\int^{+\infty}_0\,dr\,u^1_{11,2}(\kappa,r)\,\left\{
(j_0+j_2)\left(u^\prime_0(r) \right. \right.\right. \nonumber \\
&&\left.\left.\left.-u_0(r)/r\right)+[(j_0+j_2)u^\prime_2(r)+(2j_0-5j_2/2)u_2(r)/r]\st\right\} \right. \nonumber \\
&&\left. +\int^{+\infty}_0\,dr\,u^1_{11,2}(\kappa,r)\,j_1\left[-u^{\prime}_0(r)/r+u_0(r)/r^2
+\left(u^\prime_2(r)/2  \right.\right.\right. \nonumber \\
&&\left.\left.\left.+u_2(r)/r\right)/\st r \right]\right>\,.  \label{L50}
\eea
\bea
<^{3}P_1||{\hat M}_{50}||d>&=&i\frac{\st q}{2 M} [g_A F_A - \frac{g_P\,q_0}{m_\mu}]\,\int^{+\infty}_0\,dr\,u^1_{11,2}(\kappa,r)\,
j_1(u_0(r) +u_2(r)/\st)  \nonumber \\
&&-i\st \,\frac{g_A F_A}{M}\,\int^{+\infty}_0\,dr\,u^1_{11,2}(\kappa,r)\,j_0\left[u^\prime(r)-u_0(r)/r+\left(u^\prime_2(r)\right. \right. \nonumber \\
&&\left.\left. +2u_2(r)/r\right)/\st\right]\,.
\label{M50}
\eea

\subsubsection{J=1 multipoles}
\label{J1M}

\be <^{1}S_0||{\hat T}^{mag}_1||d>=-i\frac{q\, G^V_M}{\st\,
M}\int^{+\infty}_0\,dr\,u^0_{00,1}(\kappa,r)\, [j_0 u_0(r)-j_2
u_2(r)/\st]\,,  \label{T1M1S0} \ee \bea <^{1}S_0||i{\hat
T}^{el}_{51}||d>&=&-i\st g_A
F_A(1-\frac{\vq^{\,\,2}}{8M^2})\int^{+\infty}_0\,dr\,
u^0_{00,1}(\kappa,r)\,\left[j_0 u_0(r) \right.\nonumber \\
&&\left. -j_2 u_2(r)/\st\right]+i\frac{g_A F_A}{ M^2} \left\{
\frac{q}{\st}\int^{+\infty}_0\,dr\,u^0_{00,1}(\kappa,r)\,  \right.   \nonumber \\
&&\left.  j_1[u^\prime_0(r)-u_0(r)/r]
+\frac{q}{20}\int^{+\infty}_0\,dr\,u^0_{00,1}(\kappa,r)
\left[\left(7j_1  \right.\right. \right. \nonumber \\
&&\left.\left.\left. -3j_3\right)u^\prime_2(r)+(14j_1+9j_3)u_2(r)/r\right]
\right. \nonumber \\
&&\left. +\frac{1}{3 \st}\int^{+\infty}_0\,dr\,u^0_{00,1}(\kappa,r)\left[(-2j_0+j_2)u^{\prime\prime}_0(r)
+3j_2\left(-u^\prime_0(r)/r   \right.\right.\right.  \nonumber \\
&&\left.\left.\left. +u_0(r)/r^2\right)\right]-\frac{1}{3}\int^{+\infty}_0\,dr\,u^0_{00,1}(\kappa,r)
\left[(j_0-j_2/2)u^{\prime\prime}_2(r) \right.\right. \nonumber \\
&&\left.\left. +3j_0u^\prime_2(r)/r+3j_2 u_2(r)/r^2 \right]
\right\}\,,  \label{T1E1S0} \eea \bea <^{1}S_0||{\hat
L}_{\,51}||d>&=&-i[g_A
F_A(1-\frac{\vq^{\,\,2}}{8M^2})-\frac{g_P\,\vq^{\,\,2}}{2M\,m_\mu}]\,\int^{+\infty}_0\,dr\,
u^0_{00,1}(\kappa,r)\,\left[j_0 u_0(r) \right.\nonumber \\
&&\left. +\st j_2 u_2(r)\right]+i\frac{g_A F_A}{2 M^2} \left\{
\frac{3q}{5\st}\int^{+\infty}_0\,dr\,u^0_{00,1}(\kappa,r)\,  \right.   \nonumber \\
&&\left. [(j_1+j_3)u^\prime_2(r)+(2j_1-3j_3)u_2(r)/r]
\right.  \nonumber \\
&&\left. -\frac{2}{3}\int^{+\infty}_0\,dr\,u^0_{00,1}(\kappa,r)\left[(j_0+j_2)u^{\prime\prime}_0(r)
+3j_2\left(-u^\prime_0(r)/r   \right.\right.\right.  \nonumber \\
&&\left.\left.\left. +u_0(r)/r^2\right)\right]-\frac{\st}{3}\int^{+\infty}_0\,dr\,u^0_{00,1}(\kappa,r)
\left[(j_0+j_2)u^{\prime\prime}_2(r) \right.\right. \nonumber \\
&&\left.\left. +3j_0u^\prime_2(r)/r-6j_2 u_2(r)/r^2 \right]
\right\}\,,  \label{L11S0}
\eea
\bea
<^{1}S_0||{\hat M}_{51}||d>&=&i\frac{q}{2 M}(g_A F_A-\frac{g_P\, q_0}{m_\mu})\,\int^{+\infty}_0\,dr\,
u^0_{00,1}(\kappa,r)\,\left[j_0 u_0(r)+\st j_2 u_2(r)\right]  \nonumber \\
&& +i\frac{g_A F_A}{ M}\int^{+\infty}_0\,dr\,u^0_{00,1}(\kappa,r)\,j_1\,\left[u^\prime_0(r)-u_0(r)/r-\st\left(u^\prime_2(r)
\right.\right.\nonumber \\
&&\left.\left. +2u_2(r)/r \right)\right]\,,  \label{M11S0}
\eea
\bea
<^{3}P_0||{\hat T}^{mag}_{51}||d>&=&\st g_A F_AI^1_1\,,\quad
<^{3}P_0||i{\hat T}^{el}_{1}||d>=\frac{\st}{ M}[\frac{q\, G^V_M}{2}I^1_1-\frac{F^V_1}{3}K^1_1]\,,
\label{TM513P0} \\
<^{3}P_0||{\hat M}_{1}||d>&=&-F^V_1J^1_1\,,\quad
<^{3}P_0||{\hat L}_{1}||d>=\frac{F^V_1}{ M}[\frac{q}{2}J^1_1-\frac{1}{3}K^1_2]\,,
\label{M13P0}  \\
<^{3}P_1||{\hat T}^{mag}_{51}||d>&=& -\sqrt{\frac{3}{2}}g_A F_AI^1_2\,,\quad
<^{3}P_1||i{\hat T}^{el}_{1}||d>=\frac{1}{ M}\left[-\sqrt{\frac{3}{2}}\frac{q\, G^V_M}{2}I^1_2 \right. \nonumber \\
&&\left.+\sqrt{\frac{2}{3}}F^V_1K^1_3\right]\,,\quad
<^{3}P_1||{\hat M}_{1}||d>=\sqrt{3}F^V_1J^1_2\,, \label{TM513P1}
\eea
\bea
<^{3}P_1||{\hat L}_{1}||d>&=&\sqrt{3}\frac{F^V_1}{ M}\left[-\frac{q}{2}J^1_2+\frac{1}{3}K^1_4\right]\,,\,
<^{3}P_2||{\hat T}^{mag}_{51}||d>=-\sqrt{\frac{5}{2}}g_A F_AI^1_{3\lambda}\,,
\label{L13P1}  \\
<^{3}P_2||i{\hat T}^{el}_{1}||d>&=&-\sqrt{\frac{5}{2}}\frac{1}{M}\left[\frac{q\, G^V_M}{2}I^1_{3\lambda}
+\frac{2}{3}F^V_1 K^1_{5\lambda}\right]\,,
\label{TE513P2} \\
<^{3}P_2||{\hat M}_{1}||d>&=&-\sqrt{5}F^V_1J^1_{3\lambda}\,,\quad
<^{3}P_2||{\hat L}_{1}||d>=\frac{\sqrt{5}}{M}F^V_1\left[\frac{q}{2}J^1_{3\lambda}
-\frac{1}{3}K^1_{6\lambda}\right]\label{M13P2}\,,  \\
<^{3}F_2||{\hat T}^{mag}_{51}||d>&=&3\sqrt{\frac{3}{10}}g_A F_AI^1_{4\lambda}\,,\quad
<^{3}F_2||i{\hat T}^{el}_{1}||d>=\sqrt{\frac{3}{10}}\frac{1}{M}\left[\frac{3q}{2}G^V_MI^1_{4\lambda}
\right. \nonumber \\
&& \left. -2 F^V_1 K^1_{7\lambda}\right]\,,\quad
<^{3}F_2||{\hat M}_{1}||d>=-3\sqrt{\frac{3}{5}}F^V_1J^1_{4\lambda}\,,
\label{TM513F2} \\
<^{3}F_2||{\hat L}_{1}||d>&=&\sqrt{\frac{3}{5}}\frac{F^V_1}{M}\left[\frac{3q}{2}J^1_{4\lambda}-K^1_{8\lambda}\right]\,,
\label{M13F2}  \\
<^{1}D_2||i{\hat T}^{el}_{51}||d>&=&ig_A F_AI^1_5\,,\quad
<^{1}D_2||{\hat T}^{mag}_{1}||d>=i\frac{q\, G^V_M}{2 M}I^1_5\,,
\label{TE511D2}  \\
<^{1}D_2||{\hat M}_{51}||d>&=&i\frac{ q}{\st M}( g_A F_A-\frac{q_0 g_P}{m_\mu} )I^1_6
-i\st\frac{g_A F_A}{ M}I^1_7\,,  \label{M511D2} \\
<^{1}D_2||{\hat L}_{51}||d>&=&-i\st (g_A F_A-\frac{\vq^{\,\,2}
g_P}{2M m_\mu})I^1_6\,, \eea \bea
I^1_1&=&\int^{+\infty}_0\,dr\,u^0_{11,2}(\kappa,r)\,j_1\,[u_0(r)+u_2(r)/\st]\,,  \label{I11} \\
I^1_2&=&\int^{+\infty}_0\,dr\,u^1_{11,2}(\kappa,r)\,j_1\,[u_0(r)-\st u_2(r)]\,,  \label{I12} \\
I^1_{3\lambda}&=&\int^{+\infty}_0\,dr\,u^2_{11,\lambda}(\kappa,r)\,j_1\,[u_0(r)-\frac{2\st}{5}u_2(r)]\,,  \label{I13L} \\
I^1_{4\lambda}&=&\int^{+\infty}_0\,dr\,u^2_{31,\lambda}(\kappa,r)\,j_1\,u_2(r)\,,  \label{I14L}  \\
I^1_5&=&\int^{+\infty}_0\,dr\,u^2_{20,1}(\kappa,r)\,[j_2 u_0(r)-(2j_0+j_2)u_2(r)/\st]\,,  \label{I15} \\
I^1_6&=&\int^{+\infty}_0\,dr\,u^2_{20,1}(\kappa,r)\,[j_2 u_0(r)+(j_0-j_2)u_2(r)/\st]\,,  \label{I16} \\
I^1_7&=&\int^{+\infty}_0\,dr\,u^2_{20,1}(\kappa,r)\,j_1\,\left[D^1_+ u_0(r)-\st D^1_- u_2(r)\right]\,,  \label{I17}
\eea
\bea
J^1_1&=&\int^{+\infty}_0\,dr\,u^0_{11,2}(\kappa,r)\,j_1\,[u_0(r)-\st u_2(r)]\,,  \label{J11} \\
J^1_2&=&\int^{+\infty}_0\,dr\,u^1_{11,2}(\kappa,r)\,j_1\,[u_0(r)+u_2(r)/\st]\,,  \label{J12} \\
J^1_{3\lambda}&=&\int^{+\infty}_0\,dr\,u^2_{11,\lambda}(\kappa,r)\,j_1\,[u_0(r)-\frac{1}{5\st}u_2(r)]\,,  \label{J13L} \\
J^1_{4\lambda}&=&\int^{+\infty}_0\,dr\,u^2_{31,\lambda}(\kappa,r)\,j_1\,u_2(r)\,,  \label{J14L}
\eea
\bea
K^1_1&=&\int^{+\infty}_0\, dr\, u^0_{11,2}(\kappa,r)\,\left\{(j_0+j_2)D^1_+ u_0(r)
-\frac{1}{5\st}\left[(10 j_0+j_2) D^1_-  \right.\right. \nonumber \\
&&\left.\left. +9 j_2 D^3_+\right] u_2(r)\right\}\, ,  \label{K11} \\
K^1_2&=&\int^{+\infty}_0 \,dr\, u^0_{11,2}(\kappa,r)\,\left\{(j_0-2j_2)D^1_+ u_0(r)
-\frac{\st}{5}\left[(5 j_0-j_2) D^1_-  \right.\right. \nonumber \\
&&\left.\left. -9 j_2 D^3_+\right] u_2(r)\right\} \,,  \label{K12}  \\
K^1_3&=&\int^{+\infty}_0 \,dr\, u^1_{11,2}(\kappa,r)\,\left\{(j_0+j_2)D^1_+ u_0(r)
+\frac{1}{10\st}\left[(10 j_0+j_2) D^1_-  \right.\right. \nonumber \\
&&\left.\left. +9 j_2 D^3_+\right] u_2(r)\right\} \,,  \label{K13} \\
K^1_4&=&\int^{+\infty}_0\, dr\, u^1_{11,2}(\kappa,r)\,\left\{(j_0-2j_2)D^1_+ u_0(r)
+\frac{1}{5\st}\left[(5 j_0-j_2) D^1_-  \right.\right. \nonumber \\
&&\left.\left. -9 j_2 D^3_+\right] u_2(r)\right\}\, ,  \label{K14}  \\
K^1_{5\lambda}&=&\int^{+\infty}_0 \,dr\, u^2_{11,\lambda}(\kappa,r)\,\left\{(j_0+j_2)D^1_+ u_0(r)
-\frac{1}{50\st}\left[(10 j_0+j_2) D^1_-  \right.\right. \nonumber \\
&&\left.\left. +9 j_2 D^3_+\right] u_2(r)\right\} \,,  \label{K15L} \\
K^1_{6\lambda}&=&\int^{+\infty}_0\, dr\, u^2_{11,\lambda}(\kappa,r)\,\left\{(j_0-2j_2)D^1_+ u_0(r)
-\frac{1}{25\st}\left[(5 j_0-j_2) D^1_-  \right.\right. \nonumber \\
&&\left.\left. -9 j_2 D^3_+\right] u_2(r)\right\}\, ,  \label{K16L}  \\
K^1_{7\lambda}&=&\int^{+\infty}_0 \,dr\, u^2_{31,\lambda}(\kappa,r)\,\left[j_0 D^3_+
+j_2 \frac{d}{dr}\right]u_2(r)\,,\label{K17L}  \\
K^1_{8\lambda}&=&\int^{+\infty}_0 \,dr\, u^2_{31,\lambda}(\kappa,r)\,\left[j_0 D^3_+
-2j_2 \frac{d}{dr}\right]u_2(r)\,,\label{K18L}  \\
D^1_+&=&\frac{d}{dr}-\frac{1}{r}\,,\quad D^1_-=\frac{d}{dr}+\frac{2}{r}\,,\quad
D^3_+=\frac{d}{dr}-\frac{3}{r}\,. \label{DS}
\eea

\subsubsection{J=2 multipoles}
\label{J2M}

\bea
<^{1}D_2||i{\hat T}^{el}_2||d>&=&-\sqrt{5}\frac{q\,G^V_M}{2M}I^2_1\,,\quad
<^{1}D_2||{\hat T}^{mag}_{52}||d>=-\sqrt{5}g_A F_A I^2_1\,,
\label{TE21D2} \\
<^{3}P_1||i{\hat T}^{el}_{52}||d>&=&i\sqrt{\frac{3}{2}}g_A F_A I^2_2\,,\quad
<^{3}P_1||{\hat T}^{mag}_{2}||d>=i\frac{3^{3/2}}{2M}[\frac{q\,G^V_M}{3\st} I^2_2+ F^V_1 I^2_{11}],\,
\label{TE523P1} \\
<^{3}P_1||{\hat L}_{52}||d>&=&i(g_A F_A-\frac{\vq^{\,\,2} g_P}{2M
m_\mu}) I^2_5\,,\,
<^{3}P_1||{\hat M}_{52}||d>=-i\frac{q}{2M} (g_A F_A-\frac{q_0 g_P}{m_\mu})I^2_5 \nonumber\\
&& -i\frac{g_A F_A}{M}I^2_6\,,\quad
<^{3}P_2||i{\hat T}^{el}_{52}||d>=-i\frac{3}{\st} g_A F_A I^2_{3\lambda}\,, \label{EL523P1} \\
<^{3}P_2||{\hat L}_{52}||d>&=&-i\sqrt{3} \left(g_A
F_A-\frac{\vq^{\,\,2} g_P}{2M m_\mu}\right) I^2_{7\lambda}\,,\,
<^{3}P_2||{\hat T}^{mag}_{2}||d>=i\frac{3}{2M} \nonumber \\
&& \left[-\frac{q}{\st}G^V_M I^2_{3\lambda}  +F^V_1 I^2_{12\lambda}\right]\,,
\label{TM23P2}  \\
<^{3}P_2||{\hat M}_{52}||d>&=&i\sqrt{3}\frac{q}{2M} (g_A F_A-\frac{q_0 g_P}{m_\mu})I^2_{7\lambda}
+i\sqrt{3}\frac{g_A F_A}{M}I^2_{9\lambda}\,,\label{M523P2} \\
<^{3}F_2||i{\hat T}^{el}_{52}||d>&=&-i\frac{2}{\sqrt{3}} g_A F_A I^2_{4\lambda}\,,
<^{3}F_2||{\hat T}^{mag}_{2}||d>=-i\frac{1}{\sqrt{3} M}\left[q\,G^V_M I^2_{4\lambda} \right. \nonumber \\
&&\left.+6\st F^V_1 I^2_{13\lambda}\right]\,,
\label{TE523F2} \\
<^{3}F_2||{\hat L}_{52}||d>&=&i\st (g_A F_A-\frac{\vq^{\,\,2}
g_P}{2M m_\mu}) I^2_{8\lambda}\,,\,
<^{3}F_2||{\hat M}_{52}||d>=-i\frac{q}{\st M} \left[g_A F_A \right. \nonumber\\
&&\left.-\frac{q_0 g_P}{m_\mu}\right]I^2_{8\lambda}+i\st\frac{g_A F_A}{M}I^2_{10\lambda}\,,  \label{L523F2}
\eea
\bea
I^2_1&=&\int^{+\infty}_0\,dr\,u^2_{20,1}(\kappa,r) j_2 [u_0(r)+u_2(r)/\st]\,, \label{I21} \\
I^2_2&=&\int^{+\infty}_0\,dr\,u^1_{11,2}(\kappa,r) [ j_1 u_0(r)+\st(-2j_1
+3j_3)u_2(r)/5]\,, \label{I22} \\
I^2_{3\lambda}&=&\int^{+\infty}_0\,dr\,u^2_{11,\lambda}(\kappa,r) [ j_1 u_0(r)+\st(j_1
+j_3)u_2(r)/5]\,, \label{I23L} \\
I^2_{4,\lambda}&=&\int^{+\infty}_0\,dr\,u^2_{31,\lambda}(\kappa,r) [ j_3 u_0(r)+(-9j_1
+16j_3)u_2(r)/10\st]\,, \label{I24L} \\
I^2_5&=&\int^{+\infty}_0\,dr\,u^1_{11,2}(\kappa,r) [ j_1 u_0(r)-(4j_1
+9j_3)u_2(r)/5\st]\,, \label{I25} \\
I^2_6&=&\int^{+\infty}_0\,dr\,u^1_{11,2}(\kappa,r)\,j_2\, [ D^1_+ u_0(r)+D^1_- u_2(r)/\st]\,, \label{I26} \\
I^2_{7\lambda}&=&\int^{+\infty}_0\,dr\,u^2_{11,\lambda}(\kappa,r) [ j_1 u_0(r)+\st(j_1
-\frac{3}{2}j_3)u_2(r)/5]\,, \label{I27L} \\
I^2_{8\lambda}&=&\int^{+\infty}_0\,dr\,u^2_{31,\lambda}(\kappa,r) [ j_3 u_0(r)+(3j_1
+8j_3)u_2(r)/5\st]\,, \label{I28L}  \\
I^2_{9\lambda}&=&\int^{+\infty}_0\,dr\,u^2_{11,\lambda}(\kappa,r)\,j_2\, [ D^1_+ u_0(r)+D^1_- u_2(r)/\st]\,, \label{I29L} \\
I^2_{10\lambda}&=&\int^{+\infty}_0\,dr\,u^2_{31,\lambda}(\kappa,r)\,j_2\, [ D^1_+ u_0(r)+D^1_- u_2(r)/\st]\,, \label{I210L} \\
I^2_{11}&=&\int^{+\infty}_0\,dr\,u^1_{11,2}(\kappa,r)\,j_2\,u_2(r)/r\,, \label{I211} \\
I^2_{12\lambda}&=&\int^{+\infty}_0\,dr\,u^2_{11,\lambda}(\kappa,r)\,j_2\,u_2(r)/r\,, \label{I212L} \\
I^2_{13\lambda}&=&\int^{+\infty}_0\,dr\,u^2_{31,\lambda}(\kappa,r)\,j_2\,u_2(r)/r\,. \label{I213L}
\eea

\subsubsection{J=3 multipoles}
\label{J3M}

\bea
<^{3}P_2||{\hat T}^{mag}_{53}||d>&=&-\frac{6}{\sqrt{5}} g_A F_A I^3_{1\lambda}\,,\quad
<^{3}P_2||i{\hat T}^{el}_{3}||d>=-\frac{6}{\sqrt{5}M}\left[\frac{q\,G^V_M}{2}I^3_{1\lambda}
\right. \nonumber \\&&\left.+\frac{F^V_1}{35}I^3_{6\lambda}\right]\,,
<^{3}P_2||{\hat L}_{3}||d>=-\frac{3\sqrt{3}}{35\sqrt{5}M}F^V_1 I^3_{7\lambda}\,,\label{TM533P2} \\
<^{3}F_2||{\hat T}^{mag}_{53}||d>&=&2\sqrt{\frac{5}{3}} g_A F_A I^3_{2\lambda}\,,
<^{3}F_2||i{\hat T}^{el}_{3}||d>=\frac{\sqrt{5}}{M}\left[\frac{q}{\sqrt{3}}\,G^V_M I^3_{2\lambda} \right. \nonumber\\
&&\left.-\frac{2}{7}\,F^V_1 I^3_{8\lambda}\right]\,,\quad
<^{3}F_2||{\hat L}_{3}||d>=-\frac{3\sqrt{5}}{7 M}F^V_1 I^3_{9\lambda}\,,
\label{TM533F2} \\
<^{1}D_2||{\hat T}^{mag}_{3}||d>&=&-2i\frac{q\,G^V_M}{2M} I^3_3\,,\quad
<^{1}D_2||i{\hat T}^{el}_{53}||d>=-2i g_A F_A  I^3_3\,,
\label{TM31D} \\
<^{1}D_2||{\hat L}_{53}||d>&=&-i \sqrt{3}[ g_A F_A
-\frac{\vq^{\,\,2} g_P}{2M m_\mu}] I^3_4\,,\quad
<^{1}D_2||{\hat M}_{53}||d>=i \frac{\sqrt{3}q}{2M}\left[ g_A F_A \right. \nonumber \\
&&\left. -\frac{q_0 g_P}{m_\mu}\right]I^3_4 +\sqrt{3}\frac{g_A F_A}{M}I^3_5\,,
\label{L531D2}
\eea
\bea
I^3_{1\lambda}&=&\int^{+\infty}_0\,dr\,u^2_{11,\lambda}(\kappa,r) \,j_3 u_2(r)\,,  \label{I31L} \\
I^3_{2\lambda}&=&\int^{+\infty}_0\,dr\,u^2_{31,\lambda}(\kappa,r) \,j_3 [u_0(r)-u_2(r)/5\st]\,,  \label{I32L} \\
I^3_3&=&\int^{+\infty}_0\,dr\,u^2_{20,1}(\kappa,r) [j_2 u_0(r)-\st(j_2+\frac{9}{2}j_4)u_2(r)/7]\,, \label{I33} \\
I^3_4&=&\int^{+\infty}_0\,dr\,u^2_{20,1}(\kappa,r) [j_2 u_0(r)-\st(j_2-6j_4)u_2(r)/7]\,, \label{I34} \\
I^3_5&=&\int^{+\infty}_0\,dr\,u^2_{20,1}(\kappa,r) j_3 [D^1_+u_0(r)-\st D^1_- u_2(r)]\,, \label{I35} \\
I^3_{6\lambda}&=&\int^{+\infty}_0\,dr\,u^2_{11,\lambda}(\kappa,r) \,[14 j_2 D^1_- +(j_2+15 j_4)D^3_+]u_2(r)\,, \label{I36L} \\
I^3_{7\lambda}&=&\int^{+\infty}_0\,dr\,u^2_{11,\lambda}(\kappa,r) \,[14 j_2 D^1_- +(j_2-20 j_4)D^3_+]u_2(r)\,, \label{I37L} \\
I^3_{8\lambda}&=&\int^{+\infty}_0\,dr\,u^2_{31,\lambda}(\kappa,r) \,\left\{(j_2+j_4)D^1_+ u_0(r)
-\st\frac{4}{25}\left[(j_2+\frac{15}{8}j_4)D^3_+ \right.\right. \nonumber \\
&& \left.\left. +\frac{1}{2}(3j_2+\frac{5}{4}j_4)D^1_-\right]u_2(r)\right\}\,, \label{I38L}  \\
I^3_{9\lambda}&=&\int^{+\infty}_0\,dr\,u^2_{31,\lambda}(\kappa,r) \,\left\{(j_2-\frac{4}{3}j_4)D^1_+ u_0(r)
-\st\frac{4}{25}\left[(j_2-\frac{5}{2}j_4)D^3_+ \right.\right. \nonumber \\
&& \left.\left. +\frac{1}{2}(3j_2-\frac{5}{3}j_4)D^1_-\right]u_2(r)\right\}\,, \label{I39L}
\eea

We now present the multipoles J=1 of the TAA MECs given in
Appendices \ref{taawec} and \ref{fttaawec}.

\subsection{Multipoles J=1 of the TAA MECs}
\label{MTAAC}

First follow the multipoles of the weak vector MECs.
\begin{enumerate}
\item The $\pi$-pair term ${\vec {\tilde j}}^{\,\,a}(p.t.)$, Eq.\,(\ref{FTVPPT}):
\be
<^{1}S_0||{\hat T}^{mag}_{1}||d>\,=\,-i\frac{\st}{\pi}\,f^2_{\pi NN}\,F^V_1\,
\int^{+\infty}_0\,dr\,u^0_{00,1}(\kappa,r)\,j_1\,W_{1\pi}[u_0(r)+u_2(r)/\st]\,.
\label{TMVPPT}
\ee
\item The pion-in-flight term ${\vec{\tilde j}}^{\,\,a}(\pi\pi)$, Eq.\,(\ref{FTVPFT}):
\bea
<^{1}S_0||{\hat T}^{mag}_{1}||d>\,&=&\,i\frac{\sqrt{2}}{3\pi^2}\left(\frac{f_{\pi NN}}{m_\pi}\right)^2\,F^V_1
\,\int^{+\infty}_0\,dr\,u^0_{00,1}(\kappa,r)\,\left\{\left[j_0( F^2_{00}-F^2_{20})\right.\right.\nonumber \\
&&\left.\left.\,+j_2(F^2_{22}- F^2_{02})\right]u_0(r) -\left[j_2( F^2_{00}-F^2_{20}) \right.\right. \nonumber \\
&& \left.\left.\,+(j_0+2j_2)(F^2_{02}-F^2_{22})\right]u_2(r)/\st\right\} \label{TMVPFT}
\eea
\item The $\Delta$ excitation current of the $\pi$ range ${\vec {\tilde j}}^{\,\,a}_{\pi}(\Delta)$, Eq.\,(\ref{FTVDCPIR}):
\bea
<^{1}S_0||{\hat T}^{mag}_{1}||d>\,&=&\,i\frac{\st q C^V_\pi m^3_\pi F^V_1}{9 \pi (M_\Delta-M)}
\left\{\int^{+\infty}_0\,dr\,u^0_{00,1}(\kappa,r)\,W_{2\pi} \right. \nonumber \\
&& \left. \,\times \left[j_2 u_0(r)-(j_0+j_2/2)\st\,u_2(r)\right]
\right. \nonumber \\
&& \left. -f(Y,Z)\int^{+\infty}_0\,dr\,u^0_{00,1}(\kappa,r)\,(W_{2\pi}-3W_{1\pi}/x_\pi) \right. \nonumber \\
&& \left. \times [-2j_0 u_0(r)  +\st j_2 u_2(r)]\right\}\,,
\label{TMVDCPIR} \eea
\item The $\Delta$ excitation current of the $\rho$ range ${\vec {\tilde j}}^{\,\,a}_{\rho}(\Delta)$, Eq.\,(\ref{FTVDCRHOR}):
\bea <^{1}S_0||{\hat T}^{mag}_{1}||d>\,&=&\,i\frac{\st q C^V_\rho
m^3_\rho F^V_1}{9 \pi (M_\Delta-M)}
\left\{\int^{+\infty}_0\,dr\,u^0_{00,1}(\kappa,r)\,W_{2\rho} \right. \nonumber \\
&& \left. \,\times \left[j_2 u_0(r)-(j_0+j_2/2)\st\,u_2(r)\right] \right. \nonumber \\
&& \left. \,+8f(Y,Y)\int^{+\infty}_0\,dr\,u^0_{00,1}(\kappa,r)\,(W_{2\rho}-3W_{1\rho}/x_\rho) \right. \nonumber \\
&& \left. \,\times [j_0 u_0(r)  - j_2 u_2(r)/\st]\right\}\,,  \label{TMVDCRHOR}
\eea

\end{enumerate}

Now we write down the multipoles of the weak axial MECs,
\begin{enumerate}
\item
The $\pi$ potential current ${\vec{\tilde j}}^{\,\,a}_{5\pi}(p.c.)$, Eq.\,(\ref{FTAPIPT}):
\bea
<^{1}S_0||i{\hat T}^{el}_{51}||d>&=&i\frac{f^2_{\pi NN}}{2\st \pi}\frac{m_\pi}{M}g_A\,F_A\,
\int^{+\infty}_0\,dr\,u^0_{00,1}(\kappa,r)\,\left<\left[\left(qj_1 \right.\right.\right. \nonumber \\
&&\left.\left.\left. -2j_2D^1_+\right)W_{1\pi}/m_\pi+j_2W_{2\pi}\right.]u_0(r)+\left\{\left[
qj_1/2 \right.\right.\right. \nonumber \\
&&\left.\left.\left. +2j_0 D^1_- +j_2(2D^1_- +3D^3_+)/5\right]W_{1\pi}/m_\pi \right.\right. \nonumber \\
&&\left.\left. -(2j_0+j_2)W_{2\pi}/2\right\}\st\,u_2(r)\right>\,, \label{TEPIPT}
\eea
\bea
<^{1}S_0||{\hat L}_{51}||d>&=&i\frac{f^2_{\pi NN}}{4 \pi}\frac{m_\pi}{M}g_A\,F_A\,
\int^{+\infty}_0\,dr\,u^0_{00,1}(\kappa,r)\,\left<\left[\left(qj_1 \right.\right.\right. \nonumber \\
&&\left.\left.\left. +4j_2D^1_+\right)W_{1\pi}/m_\pi-2j_2W_{2\pi}\right.]u_0(r)+\left\{\left[
-qj_1 \right.\right.\right. \nonumber \\
&&\left.\left.\left. +2j_0 D^1_- -2j_2(2D^1_- +3D^3_+)/5\right]W_{1\pi}/m_\pi \right.\right. \nonumber \\
&&\left.\left. +(j_2-j_0)W_{2\pi}\right\}\st\,u_2(r)\right>\,, \label{LPIPT}
\eea
\item The $\rho$-$\pi$ current ${\vec{\tilde j}}^{\,\,a}_{5}(\rho\pi)$, Eq.\,(\ref{FTRHPIT}):
\bea
<^{1}S_0||i{\hat T}^{el}_{51}||d>&=&i\bigg(\frac{f_{\pi NN}}{ m_\pi}\bigg)^2\,\frac{(1+\kappa^V_\rho)m^2_\rho}
{6\st \pi^2 Mg_A q}\,e^{-q^2(1-a)/2\Lambda^2_\rho}[2{\bar I}^0_1+{\bar I}^2_1]\,, \label{TERHPI} \\
<^{1}S_0||{\hat L}_{51}||d>&=&i\bigg(\frac{f_{\pi NN}}{ m_\pi}\bigg)^2\,\frac{(1+\kappa^V_\rho)m^2_\rho}
{6 \pi^2 Mg_A q}\,e^{-q^2(1-a)/2\Lambda^2_\rho}[{\bar I}^0_1-{\bar I}^2_1]\,, \label{LRHPI}
\eea
\bea
{\bar I}^0_1&=&\int^{+\infty}_0 dr u^0_{00,1}(\kappa,r)j_0(bqr)[(H^2_{00}-H^2_{20}) u_0(r)
+(H^2_{22}-H^2_{02})u_2(r)/\st]\,,  \label{BI01} \\
{\bar I}^2_1&=&\int^{+\infty}_0\,dr\,u^0_{00,1}(\kappa,r)\,j_2(bqr)\,\left[(H^2_{02}-H^2_{22}) u_0(r)
+\left(-2H^2_{00}-H^2_{02} \right.\right. \nonumber \\
&&\left.\left. +H^2_{20}/5+16H^2_{22}/7+108 H^2_{24}/35\right)u_2(r)/\st\right]\,, \label{BI21} \\
b&=&1/2\,-\,a\,.  \label{b}
\eea
These multipole contributions correspond to the part proportional to
$\Delta_1 {\vec{\tilde j}}^{\,\,a}_{5}(\rho\pi)$ of the current (\ref{FTRHPIT}). The contributions
due to the part $\Delta_2 {\vec{\tilde j}}^{\,\,a}_{5}(\rho\pi)$ can be obtained
from Eqs.\,(\ref{TERHPI}) and (\ref{LRHPI}) by the change (\ref{CHANGE}).
\item The $\Delta$ excitation current of the $\pi$ range ${\vec {\tilde j}}^{\,\,a}_{5\pi}(\Delta)$, Eq.\,(\ref{FTADCPIR}):
\bea
<^{1}S_0||i{\hat T}^{el}_{51}||d>&=&i\frac{\st g_A F_A m^3_\pi}{9 \pi(M_\Delta-M)}
\bigg(\frac{f_{\pi N \Delta}}{m_\pi}\bigg)^2\,(m^2_\pi-q^2_0)\Delta^\pi_F(q^2) \nonumber \\
&& \times \left\{\int^{+\infty}_0\,dr\,u^0_{00,1}(\kappa,r)\,W_{2\pi}[j_2 u_0(r)-(j_0+j_2/2)\st\,u_2(r)]
 \right. \nonumber \\
&& \left. +\frac{1}{2}f(Z,Z)\int^{+\infty}_0\,dr\,u^0_{00,1}(\kappa,r)\,(W_{2\pi}-3W_{1\pi}/x_\pi)
\right. \nonumber \\ && \left. \times [-2j_0 u_0(r)+\st j_2 u_2(r)]\right\}\,,  \label{TEADCPIR} \\
<^{1}S_0||{\hat L}_{51}||d>&=&i\frac{2 g_A F_A m^3_\pi}{9 \pi(M_\Delta-M)}
\bigg(\frac{f_{\pi N \Delta}}{m_\pi}\bigg)^2\,(m^2_\pi-q^2_0)\Delta^\pi_F(q^2) \nonumber \\
&& \times \left\{\int^{+\infty}_0\,dr\,u^0_{00,1}(\kappa,r)\,W_{2\pi}[-j_2 u_0(r)+(j_2-j_0)u_2(r)/\st]
 \right. \nonumber \\
&& \left. -\frac{1}{2}f(Z,Z)\int^{+\infty}_0\,dr\,u^0_{00,1}(\kappa,r)\,(W_{2\pi}-3W_{1\pi}/x_\pi)
\right. \nonumber \\ && \left. \times [j_0 u_0(r)+\st j_2 u_2(r)]\right\}\,.  \label{LADCPIR}
\eea
\item The $\Delta$ excitation current of the $\rho$ range ${\vec {\tilde j}}^{\,\,a}_{5\rho}(\Delta)$, Eq.\,(\ref{FTADCRHOR}):
\bea
<^{1}S_0||i{\hat T}^{el}_{51}||d>&=&-i\frac{\st g_A F_A\,m^3_\rho C^A_\rho}{9\pi (M_\Delta-M)}\,
(m^2_\pi-q^2_0)\Delta^\pi_F(q^2) \nonumber \\
&& \times \left\{\int^{+\infty}_0\,dr\,u^0_{00,1}(\kappa,r)\,W_{2\rho}[j_2 u_0(r)-(j_0+j_2/2)\st\,u_2(r)]
 \right. \nonumber \\
&& \left. -4f(Y,Z)\int^{+\infty}_0\,dr\,u^0_{00,1}(\kappa,r)\,(W_{2\rho}-3W_{1\rho}/x_\rho)
\right. \nonumber \\ && \left. \times [j_0 u_0(r)-j_2 u_2(r)/\st]\right\}\,,  \label{TEADCRHOR} \\
<^{1}S_0||{\hat L}_{51}||d>&=&-i\frac{2 g_A F_A\,m^3_\rho C^A_\rho}{9\pi (M_\Delta-M)}\,(m^2_\pi-q^2_0)\Delta^\pi_F(q^2) \nonumber \\
&& \times \left\{\int^{+\infty}_0\,dr\,u^0_{00,1}(\kappa,r)\,W_{2\rho}[-j_2 u_0(r)+(j_2-j_0)u_2(r)/\st]
 \right. \nonumber \\
&& \left. -2f(Y,Z)\int^{+\infty}_0\,dr\,u^0_{00,1}(\kappa,r)\,(W_{2\rho}-3W_{1\rho}/x_\rho)
\right. \nonumber \\ && \left. \times \left[j_0 u_0(r)+\st j_2 u_2(r)\right]\right\}\,.  \label{LADCRHOR}
\eea
\item The potential current of the $\rho$ range ${\vec{\tilde j}}^{\,\,a}_{5\rho}(p.c.)$, Eq.\,(\ref{FTAPCRHOR}):
\bea
<^{1}S_0||i{\hat T}^{el}_{51}||d>&=&ig_A F_A\left(\frac{g_\rho}{2}\right)^2
\left(\frac{m_\rho}{M}\right)^3\frac{(1+\kappa^V_\rho)^2}{8\st \pi} \nonumber \\
&& \times \left<\int^{+\infty}_0\,dr\,u^0_{00,1}(\kappa,r)\left[(qj_1+2j_2 D^1_+)W_{1\rho}/m_\rho-j_2W_{2\rho}\right]u_0(r)
\right. \nonumber \\
&&\left. +\int^{+\infty}_0\,dr\,u^0_{00,1}(\kappa,r)\left\{[q(8j_1+3j_3)/5
-2(2j_0+j_2)D^1_-]W_{1\rho}/m_\rho  \right.\right.\nonumber \\
&&\left.\left. +(2j_0+j_2)W_{2\rho}\right\}u_2(r)/\st \right>\,. \label{TEAPCRHOR} \\
<^{1}S_0||{\hat L}_{51}||d>&=&-ig_A F_A\left(\frac{g_\rho}{2}\right)^2
\left(\frac{m_\rho}{M}\right)^3\frac{(1+\kappa^V_\rho)^2}{8 \pi} \nonumber \\
&& \times \left<\int^{+\infty}_0\,dr\,u^0_{00,1}(\kappa,r)\left[(qj_1+2j_2 D^1_+)W_{1\rho}/m_\rho-j_2W_{2\rho}\right]u_0(r)
\right. \nonumber \\
&&\left. +\int^{+\infty}_0\,dr\,u^0_{00,1}(\kappa,r)\left\{[q(3j_3-7j_1)/5
+2(j_0-j_2)D^1_-]W_{1\rho}/m_\rho  \right.\right.\nonumber \\
&&\left.\left. +(j_2-j_0)W_{2\rho}\right\}u_2(r)/\st \right>\,. \label{LAPCRHOR}
\eea
\item The time component of the weak axial MEC ${\tilde \rho}^a_5(\rho\pi)$, Eq.\,(\ref{FTTCAMEC}):
\bea
<^{1}S_0||{\hat M}_{51}||d>&=&i\bigg(\frac{f_{\pi NN}}{ m_\pi}\bigg)^2\,\frac{m^2_\rho}{2\pi^2 g_A}\,
e^{-q^2(1-a)/2\Lambda^2_\rho}  \nonumber \\
&& \times \left<a \int^{+\infty}_0 dr\, u^0_{00,1}(\kappa,r)\left\{[j_0(bqr) H^0_{00} +2 j_2(bqr) H^0_{22}]u_0(r)
 \right.\right. \nonumber \\
&&\left.\left. +[j_2(bqr) H^0_{00} + (j_0(bqr)-j_2(bqr))H^0_{22}]\st\,u_2(r)\right\} \right. \nonumber  \\
&&\left. +\frac{1}{q}\int^{+\infty}_0 dr\, u^0_{00,1}(\kappa,r)\,j_1(bqr)\left\{(H^1_{01}+2H^1_{21})u_0(r)
\right.\right. \nonumber \\
&&\left.\left.  -[H^1_{01}+(H^1_{21}-9
H^1_{23})/5]\st\,u_2(r)]\right\}\right>\,. \label{MTCAMEC} \eea This
multipole contribution corresponds to the part proportional to
$\Delta_1 {\tilde \rho}^{a}_{5}(\rho\pi)$ of the current
(\ref{FTTCAMEC}). The contribution due to the part $\Delta_2 {\tilde
\rho}^{a}_{5}(\rho\pi)$ can be obtained from Eq.\,(\ref{MTCAMEC}) by
the change (\ref{CHANGE}).
\end{enumerate}

Next follow the multipoles J=1 of the EFT MECs presented in Appendices
\ref{eftwec} and \ref{ftwaeftmec}.

\subsection{Multipoles J=1 of the EFT MECs}
\label{MEFTC}

\begin{enumerate}
\item The time component of the weak axial MECs, ${\tilde{\hat A}^0}_{2B}$, Eq.\,(\ref{FTA02B}),

\be
<^{1}S_0||{\hat M}_{51}||d>=-i\frac{g_A F_A m^2_\pi}{4\pi f^2_\pi}
\int^{+\infty}_0\,dr\,u^0_{00,1}(\kappa,r)\,j_1\,W_{1\pi}\,[u_0(r)-\st\, u_2(r)]\,. \label{MEFT}
\ee

\item The ${\tilde{\hat {\vec A}}}_{2B}$ term, Eq.\,(\ref{FTA2BSC}),
\bea <^{1}S_0||i{\hat T}^{el}_{51}||d>&=&i\frac{g_A F_A m^3_\pi}{\st \pi M f^2_\pi}
\int^{+\infty}_0\,dr\,u^0_{00,1}(\kappa,r)\,\left<\left\{-(3/4+2\hat c_4+\hat c_3)j_0 W_{1\pi}/x_\pi
\right.\right. \nonumber \\
&&\left.\left. +[(3/4+2\hat c_4+\hat c_3)j_0 +(\hat c_3 -\hat c_4)j_2]W_{2\pi}/3
+\left[-2(j_0+j_2)D^1_+/3   \right.\right.\right. \nonumber \\
&& \left.\left.\left. +(1+c_6)qj_1\right]W_{1\pi}/4m_\pi -(\hat d_1+2\hat d_2)j_0 W_\pi\right\}u_0(r)
\right. \nonumber \\
&&\left. +\left\{(3/4+2\hat c_4+\hat c_3)j_2 W_{1\pi}/2 x_\pi +\left[(\hat c_4 -\hat c_3)j_0
-\left(3/8+\hat c_4/2
\right.\right.\right.\right. \nonumber \\
&&\left.\left.\left.\left. +\hat c_3\right)j_2\right]W_{2\pi}/3
+\left[2j_0 D^1_-/3 + j_2(D^1_- + 9 D^3_+ )/15  \right.\right. \right.\nonumber \\
&&\left.\left.\left. +(1+c_6)q j_1/2\right]W_{1\pi}/4m_\pi +(\hat d_1+2\hat d_2)
j_2 W_\pi/2 \right\} \right. \nonumber \\ && \left. \times \st\, u_2(r)\right>\,, \label{TEA2B}
\eea
\bea <^{1}S_0||{\hat L}_{51}||d>&=&i\frac{g_A F_A m^3_\pi}{2\pi M f^2_\pi}
\int^{+\infty}_0\,dr\,u^0_{00,1}(\kappa,r)\,\left<\left\{-(3/4+2\hat c_4+\hat c_3)j_0 W_{1\pi}/x_\pi
\right.\right. \nonumber \\
&&\left.\left. +[(3/4+2\hat c_4+\hat c_3)j_0 +2(\hat c_4 -\hat c_3)j_2]W_{2\pi}/3
+\left[2(2j_2-j_0)D^1_+/3   \right.\right.\right. \nonumber \\
&& \left.\left.\left. +qj_1\right]W_{1\pi}/4m_\pi -(\hat d_1+2\hat d_2)j_0 W_\pi\right\}u_0(r)
\right. \nonumber \\
&&\left. +\left\{-(3/4+2\hat c_4+\hat c_3)j_2 W_{1\pi}/x_\pi +\left[(\hat c_4 -\hat c_3)j_0
\right.\right.\right. \nonumber \\
&&\left.\left.\left. +(3/4+2\hat c_3+\hat c_4)j_2\right]W_{2\pi}/3
+\left[2j_0 D^1_-/3 -2j_2(D^1_- + 9 D^3_+ )/15  \right.\right. \right.\nonumber \\
&&\left.\left.\left.-q j_1/2\right]W_{1\pi}/4m_\pi -(\hat d_1+2\hat d_2)
j_2 W_\pi/2 \right\}\st\, u_2(r)\right>\,. \label{LA2B}
\eea
\item
The contribution due to the ${\tilde{\hat P}}(\qa,\qb)$ term, Eq.\,(\ref{FTHP}),
\be
<^{1}S_0||{\hat L}_{51}||d>=i\frac{g_A F_A m^4_\pi}{\pi M f^2_\pi}{\hat c}_1\,\Delta^\pi_F(\vq^{\,\,2})
\int^{+\infty}_0\,dr\,u^0_{00,1}(\kappa,r)\,j_1\,W_{1\pi}\,[u_0(r)-\st\, u_2(r)]\,. \label{LPEFT}
\ee
\end{enumerate}

\end{document}